%% file: all.tex
\documentclass[12pt,a4paper]{report}
\usepackage[cp1250]{inputenc}
\usepackage{wrapfig}

\usepackage{amsmath}
\usepackage{graphicx}
\usepackage{amssymb}
\usepackage{url}
\usepackage{longtable}

\newcommand{\lsim}{\mathrel{\mathop{\kern 0pt \rlap
  {\raise.2ex\hbox{$<$}}}
  \lower.9ex\hbox{\kern-.190em $\sim$}}}
\newcommand{\gsim}{\mathrel{\mathop{\kern 0pt \rlap
  {\raise.2ex\hbox{$>$}}}
  \lower.9ex\hbox{\kern-.190em $\sim$}}}
\newcommand{\titl}[1]{{\centering\Large\bf #1\par}\bigskip}
\newcommand{\name}[1]{{\centering\rm\normalsize #1\par}\bigskip}

\newcommand{\adr}[1]{{\it \normalsize #1\par}\medskip}

\begin{document}
\input{./title/title} 
\clearpage

\addcontentsline{toc}{section}{
{\bf Introduction}
}
\input{./intro/intro} 
\setcounter{equation}{0} 
\setcounter{figure}{0}
\clearpage

\addcontentsline{toc}{section}{
{\bf Is quantum theory exact?
Collapse Models and the possibility of a break down of quantum mechanics towards the macroscopic scale}\\
A.~Bassi}
\input{./bassi/bassi} 
\setcounter{equation}{0} 
\setcounter{figure}{0}
\clearpage

\addcontentsline{toc}{section}{
{\bf The Dark Matter annual modulation signature as a probe of the dark side 
of the Universe}\\
P.~Belli, {\it et al.}}
\input{./belli/belli} 
\setcounter{equation}{0} 
\setcounter{figure}{0}
\clearpage

\addcontentsline{toc}{section}{
{\bf Future Plans for the VIP Experiment Upgrade.}\\
A.~Clozza, {\it et al.}}
\input{./clozza/clozza} 
\setcounter{equation}{0} 
\setcounter{figure}{0}
\clearpage

\addcontentsline{toc}{section}{
{\bf From the Pauli Exclusion Principle violation tests (VIP experiment) 
to collapse models experimental investigation plans}\\
C.~Curceanu, {\it et al.}}
\input{./curceanu/curceanu} 
\setcounter{equation}{0} 
\setcounter{figure}{0}
\clearpage

\addcontentsline{toc}{section}{
{\bf Probabilistic theories: classical, quantum and beyond quantum}\\
B.~Daki{\' c}}
\input{./dakic/dakic} 
\setcounter{equation}{0} 
\setcounter{figure}{0}
\clearpage

\addcontentsline{toc}{section}{
{\bf A Quantum Digital World:\\ Quantum Fields as Quantum Automata}\\ 
G. M. D'Ariano}
\input{./dariano/dariano} 
\setcounter{equation}{0} 
\setcounter{figure}{0}
\clearpage

\addcontentsline{toc}{section}{
{\bf Quantum decoherence in nuclear collision dynamics}\\
A. Diaz-Torres} 
\input{./diaz-torres/diaz-torres} 
\setcounter{equation}{0} 
\setcounter{figure}{0}
\clearpage

\addcontentsline{toc}{section}{
{\bf $CPT$ symmetry, Quantum Mechanics, and Neutral kaons}\\
A.~Di~Domenico}
\input{./didomenico/didomenico} 
\setcounter{equation}{0} 
\setcounter{figure}{0}
\clearpage

\addcontentsline{toc}{section}{
{\bf Non-Markovian features in quantum dynamics}\\
L.~Ferialdi}
\input{./ferialdi/ferialdi} 
\setcounter{equation}{0} 
\setcounter{figure}{0}
\clearpage

\addcontentsline{toc}{section}{
{\bf Quantum contextuality and identical particles}\\
R. Floreanini, {\it et al.}}
\input{./floreanini/floreanini}
\setcounter{equation}{0} 
\setcounter{figure}{0}
\clearpage

\addcontentsline{toc}{section}{
{\bf Study of rare processes with the Borexino detector}\\
K.~Fomenko, {\it et al.}}
\input{./fomenko/fomenko}
\setcounter{equation}{0} 
\setcounter{figure}{0}
\clearpage

\addcontentsline{toc}{section}{
{\bf Matter-wave interferometry and metrology with macromolecules}\\
S.~Gerlich {\it et al.}}
\input{./gerlich/gerlich}
\setcounter{equation}{0} 
\setcounter{figure}{0}
\clearpage

\addcontentsline{toc}{section}{
{\bf A search for the de Broglie particle internal clock by means of electron channeling}\\
M.~Gouanere}
\input{./gouanere/gouanere}
\setcounter{equation}{0} 
\setcounter{figure}{0}
\clearpage

\addcontentsline{toc}{section}{
{\bf Probability amplitudes of two-levels atoms beyond the dipole
       approximation}\\
A.~G.~Hayrapetyan, {\it et al.}}
\input{./hayrapetyan/hayrapetyan}
\setcounter{equation}{0} 
\setcounter{figure}{0}
\clearpage

\addcontentsline{toc}{section}{
{\bf Bell's [Un]Speakables in the Neutral Kaon System}\\
B.C.~Hiesmayr}
\input{./hiesmayr/hiesmayr}
\setcounter{equation}{0} 
\setcounter{figure}{0}
\clearpage

\addcontentsline{toc}{section}{
{\bf  Finite-speed causal-influence models \\for quantum theory lead to superluminal signaling}\\
Y-C. Liang, {\it et al.}}
\input{./liang/liang}
\setcounter{equation}{0} 
\setcounter{figure}{0}
\clearpage

\addcontentsline{toc}{section}{
{\bf  Further investigation on electron stability and 
non-paulian transition in NaI(Tl) crystals}\\ 
A. Di Marco, {\it et al.}}
\input{./marco/marco}
\setcounter{equation}{0} 
\setcounter{figure}{0}
\clearpage

\addcontentsline{toc}{section}{
{\bf  Probing Quantum-Gravity induced Decoherence in (Astro)Particle Physics}\\
N.~E.~Mavromatos}
\input{./mavromatos/mavromatos}
\setcounter{equation}{0} 
\setcounter{figure}{0}
\clearpage

\addcontentsline{toc}{section}{
{\bf  Randomness and Information Transfer in Quantum Measurements}\\
S.~Mayburov}
\input{./mayburov/mayburov}
\setcounter{equation}{0} 
\setcounter{figure}{0}
\clearpage

\addcontentsline{toc}{section}{
{\bf  Fuzzy Space-time Topology and Gauge Fields}\\
S.~Mayburov}
\input{./mayburov2/mayburov2}
\setcounter{equation}{0} 
\setcounter{figure}{0}
\clearpage

\addcontentsline{toc}{section}{
{\bf  INRIM recent results about QM foundations investigation}\\
F.~Piacentini}
\input{./piacentini/piacentini}
\setcounter{equation}{0} 
\setcounter{figure}{0}
\clearpage

\addcontentsline{toc}{section}{
{\bf  Hadron interferometry with neutrons}\\
H. Rauch}
\input{./rauch/rauch}
\setcounter{equation}{0} 
\setcounter{figure}{0}
\clearpage

\addcontentsline{toc}{section}{
{\bf  Nanofiber Photonics and Quantum Optics}\\
A.~Rauschenbeutel}
\input{./rauschenbeutel/rauschenbeutel}
\setcounter{equation}{0} 
\setcounter{figure}{0}
\clearpage

\addcontentsline{toc}{section}{
{\bf  Spontaneous X-ray emission by free electrons in the collapse models: a closer look}\\
A.~Rizzo, {\it et al.}}
\input{./rizzo/rizzo}
\setcounter{equation}{0} 
\setcounter{figure}{0}
\clearpage

\addcontentsline{toc}{section}{
{\bf  Non-Markovian dynamics: characterizations and measures}\\
B.~Vacchini}
\input{./vacchini/vacchini}
\setcounter{equation}{0} 
\setcounter{figure}{0}
\clearpage

\addcontentsline{toc}{section}{
{\bf  Two slits are too many for just one particle}\\
N.~Vona}
\input{./vona/vona}
\setcounter{equation}{0} 
\setcounter{figure}{0}
\clearpage

\addcontentsline{toc}{section}{
{\bf  Testing $\mathcal{CPT}$ with antiprotonic helium and antihydrogen}\\
E.~Widmann}
\input{./widmann/widmann}
\setcounter{equation}{0} 
\setcounter{figure}{0}
\clearpage

\addcontentsline{toc}{section}{
{\bf  Measuring the magnetic birefringence of vacuum: status and perspectives of the PVLAS experiment}\\
G.~Zavattini, {\it et al.}}
\input{./zavattini/zavattini}
\setcounter{equation}{0} 
\setcounter{figure}{0}
\clearpage

\addcontentsline{toc}{section}{
{\bf Program} }
\input{./program/program}
\setcounter{equation}{0} 
\setcounter{figure}{0}
\clearpage

\addcontentsline{toc}{section}{
{\bf List of participants} }
\input{./participants/participants}
\setcounter{equation}{0} 
\setcounter{figure}{0}
\clearpage

\addcontentsline{toc}{section}{
{\bf Conference Photos} }
\input{./photo/photo}
\setcounter{equation}{0} 
\setcounter{figure}{0}
\clearpage

\end{document}

%% file: title/title.tex
\begin{titlepage}
\begin{center}

\includegraphics[width=0.75\textwidth]{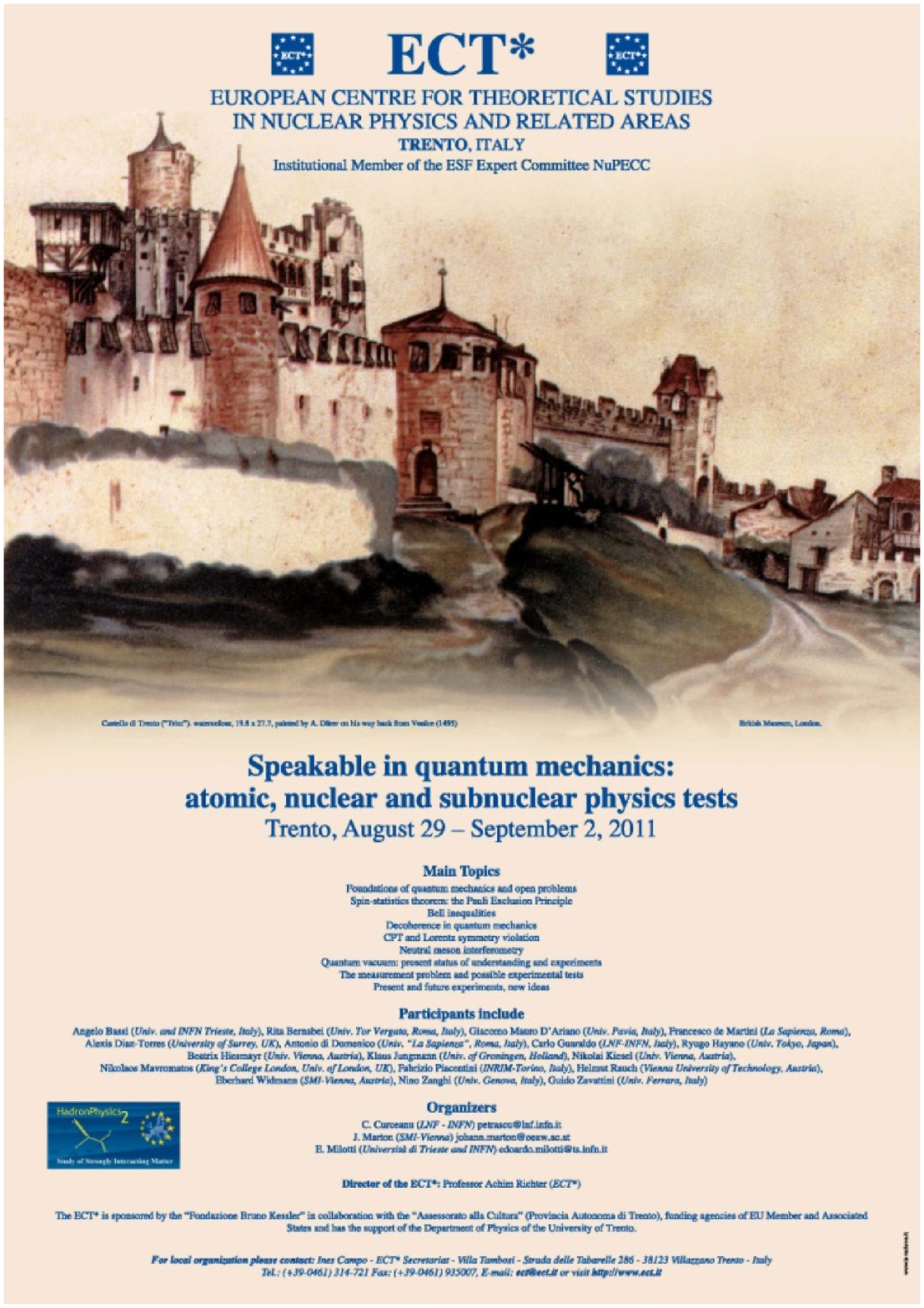}\\[1cm]
\textsc{\Large Mini-Proceedings}\\[0.5cm]
\textsc{\Large ECT* Workshop}\\[1.0cm]
\textsc{\LARGE Speakable in quantum mechanics: atomic, nuclear and subnuclear physics tests}\\[2cm]
\text{Eds.: C. Curceanu (INFN-LNF/Frascati), 
J. Marton (SMI/Vienna)}
\text{
E. Milotti (Universit{\`a} di Trieste and INFN)
}\\[1.5cm]

\end{center}
\end{titlepage}

\tableofcontents

%% file: intro/intro.tex
%






\titl{Introduction}


\titl{ ``Speakable in quantum mechanics: atomic, nuclear and subnuclear physics tests''}

Quantum Mechanics (QM) has been with us for more than 100 years by now, 
and some long-standing debates are still alive and waiting for a solution. 
It calls for creativity, as well as for exceptional skills, both in theory and experiment. 
At present, new, extremely refined methods, which allow for high precision tests of QM in the atomic,
nuclear and subnuclear fields, exist, or are around the corner.

Devoted to the hot items of QM, the international workshop ``Speakable in quantum mechanics: 
atomic, nuclear and subnuclear physics tests'', organized by Catalina Curceanu 
(LNF-INFN, Frascati, Italy), Johann Marton (SMI and TU Vienna, Austria) 
and Edoardo Milotti (Univ. and INFN Trieste, Italy), was held at the European 
Centre for Theoretical Physics (ECT*) in Trento/Italy from August 29th 
until September 2nd 2011. About 40 scientists participated to this workshop, 
with a well balanced mix of theory and experiment. Several young scientists gave interesting talks 
and contributed to the discussions -- an extremely good sign for the future development 
of the QM research field.

Many of the fascinating topics discussed in this workshop are longstanding problems that 
have been puzzling the physics community for many years. New experimental techniques and 
a new theoretical understanding - which were both well represented at the workshop -- hold the 
tantalizing promise of future breakthroughs. 

For those who want to have more information about the Workshop, you can find it  at the dedicated website:\\
\url{https://smimac0.smi.oeaw.ac.at/conference/ect_star_2011/index.html}

\vspace{3cm}
Catalina Curceanu\\
Johann Marton\\
Edoardo Milotti

\vfill 


%% file: bassi/bassi.tex

%





\titl{Is quantum theory exact?
Collapse Models and the possibility of a break down of quantum mechanics towards the macroscopic scale}

\name{
A.~Bassi$^{1,2}$
}

\adr{
$^1$ Department of Physics, University of Trieste, Strada Costiera 11, 34151 Trieste, Italy \\
$^2$  INFN, Sezione di Trieste, Via Valerio 2, 34127 Trieste,
Italy
}

Quantum mechanics is the most successful physical theory in the history of science. It set out to explain the atomic structure of matter and it has led to revolutionary technologies. Despite the successes, the debate about the precise formulation and the meaning of quantum theory is still open and very much debated. J.S. Bell stated it very clearly [1]: ``The formulations of quantum mechanics that you find in the books involve dividing the world into an observer and an observed, and you are not told where that division comes---on which side of spectacles it comes, for example, or at which end of my optic nerve [...] So you have a theory which is fundamentally ambiguous	[...]''.

Collapse models [2-4] were devised in order to resolve this abiguity. There, one assumes that the Schr\"odinger equation is not exact: A more precise description of quantum systems includes non-linear and stochastic terms in addition to the usual ones. A first general question is how to modify the Schr\"odinger equation, without violating some basic facts about Physics. It has been shown [5,6] that by requiring: i) Probability conservation and ii) No faster-than-light signaling, the form of the new equation is almost uniquely defined: 
\begin{equation}
d |\psi \rangle_t \; = \; \left[ - \frac{i}{\hbar} H dt \; + \; \sqrt{\lambda} (A - \langle A \rangle_t ) dW_t - \frac{\lambda}{2} (A - \langle A \rangle_t )^2 dt \right] |\psi\rangle_t,
\end{equation}
where $H$ is the standard quantum Hamiltonian, $A$ is the operator on whose eigenstates the wave function collapses, $\langle A \rangle_t \equiv \langle \psi_t | A | \psi_t \rangle$ is the quantum average (this is where non-linearity enters), $W_t$ is a standard Wiener process (this is where stochasticity enters), and $\lambda$ is a positive constant that sets the collapse strength.

If one chooses for $A$ the position operator, or a function of the position operator, one has {\it space} collapse models [2-4], the most widely discussed in the literature. With an appropriate choice of the collapse strength $\lambda$, these models make sure that: i) Microscopic systems behave quantum mechanically, as confirmed by experimental data; ii) Macroscopic objects behave classically, in particular they are always well-localized in space; iii) In measurement processes, the standard phenomenology (definite outcomes, the projection postulate, the Born rule) is naturally recovered. In this way, one has a consistent description of physical phenomena, at the non-relativistic level at least. 

Collapse models make predictions which differ from those of standard quantum mechanics [4]. One of the most exciting task is to perform cutting-edge experiments, in order to asses whether quantum mechanics is exact, or an approximation of a deeper level theory. \\

\vfill  

\noindent{\bf References }
\begin{description}
\setlength\itemsep{-3pt}
\item{[1]} J.S. Bell, in: {\it The Ghost in the Atom}, P.C.W. Davies ed.  (C.U.P., Cambridge, 1993).
\item{[2]} G.C. Ghirardi, A. Rimini, T. Weber, Phys. Rev. D {\bf 34} (1986) 470.
\item{[3]} A. Bassi, G.C. Ghirardi, Phys. Rept. {\bf 379} (2003) 257.
\item{[4]} S.L. Adler, A. Bassi, Science {\bf 325} (2009) 275. 
\item{[5]} S.L. Adler, T.A. Brun, J. Phys. A {\bf 34} (2001) 4797.
\item{[6]} N. Gisin, Hel. Phys. Acta {\bf 62} (1989) 363.
\end{description}


%% file: belli/belli.tex

%






\titl{The Dark Matter annual modulation signature as a probe of the dark side 
of the Universe}

\name{ 
R. Bernabei$^{1}$, P. Belli$^{1}$, F. Cappella$^{2}$, R. Cerulli$^{3}$, A. d'Angelo$^{2}$,
C. J. Dai$^{4}$, A. Di Marco$^{1}$, H.L. He$^{4}$, A. Incicchitti$^{2}$,
X. H. Ma$^{4}$, F. Montecchia$^{1,5}$, D. Prosperi$^{\dagger,2}$,X. D. Sheng$^{4}$, R.G. 
Wang$^{4}$, Z.P. Ye$^{4,6}$
}
\adr{
$^1$ Dip. di Fisica, Univ. Roma ``Tor Vergata'' and INFN, sez. Roma ``Tor Vergata'', Rome 
I-00133\\
$^2$ Dip. di Fisica, Univ. Roma ``La Sapienza'' and INFN, sez. Roma ``La Sapienza'', Rome 
I-00183\\
$^3$ Laboratori Nazionali del Gran Sasso, I.N.F.N., Assergi, Italy\\
$^4$ IHEP, Chinese Academy, P.O. Box 918/3, Beijing 100039, China\\
$^5$ Laboratorio Sperimentale Policentrico di Ingegneria Medica, Univ. Roma ``Tor 
Vergata''\\
$^6$ University of Jing Gangshan, Jiangxi, China\\
$^ \dagger$ Deceased\\
}


The DAMA/NaI and DAMA/LIBRA experiments at the Gran Sasso
underground laboratory have been and are, respectively, 
investigating the presence of the Dark Matter (DM) particles in the galactic halo
by exploiting the model independent DM annual modulation signature.
In fact, as a consequence of its annual revolution around the Sun, 
which is moving in the Galaxy, the Earth should be crossed by a larger 
flux of DM particles around $\sim$2 June 
(when the Earth orbital velocity is summed to the one of the
solar system with respect to the Galaxy) and by a smaller one 
around $\sim$2 December (when the two velocities are subtracted). 
This DM annual modulation signature 
is very distinctive and unambiguous, since the effect induced by DM particles must 
simultaneously satisfy many strong requirements.
To mimic such a signature spurious effects or side reactions should be able 
not only to account for the observed 
modulation amplitude but also to simultaneously satisfy all the requirements of the signature.

The DAMA/LIBRA set-up
is made of 25 highly radiopure 9.70 kg mass NaI(Tl) crystal scintillators.
The software energy threshold has been cautiously taken at 2 keV electron equivalent
and the light response is 5.5--7.5 photoelectrons/keV depending on the detector.
The detectors are housed in a low radioactivity sealed 
copper box installed in the center of a low-radioactivity multi-component shield.

The DAMA/LIBRA data released so far correspond to
six annual cycles for an exposure of 0.87 ton$\times$yr.
Considering these data with those of DAMA/NaI
over 7 annual cycles, the total exposure collected
is 1.17 ton$\times$yr. 
Several analyses on the model-independent DM annual
modulation signature have been performed.

The results provide a model
independent evidence of the presence of DM particles in the galactic halo at 8.9 $\sigma$ C.L.
on the basis of the investigated DM signature.
The modulation amplitude of the single-hit events in the (2--6) keV energy bin
is $(0.0116 \pm 0.0013)$
cpd/kg/keV; the phase is $(146 \pm 7)$ days and the period is $(0.999\pm0.002)$ yr, values well in agreement
with those expected for the DM particles.

Careful investigations
on any significant systematics or side reaction able to
account for the measured modulation amplitude and to simultaneously satisfy
all the requirements of the signature
have been carried out.
No systematics or side reactions able to mimic the signature (that is, able to
account for the measured modulation amplitude and simultaneously satisfy
all the requirements of the signature) has been found or suggested
over more than a decade.

\vfill 



%


%% file: clozza/clozza.tex

%





\titl{Future Plans for the VIP Experiment Upgrade.}

\name{
A.~Clozza$^{1}$

(On behalf of the VIP Collaboration)}

\adr{
$^1$ INFN, Laboratori Nazionali di Frascati, CP 13, Via E. Fermi 40, I-00044 Frascati (Roma), Italy \\
}


The Pauli Exclusion Principle (PEP) is a consequence of the spin-statistics connection [1] and plays a fundamental role in our understanding of many physical and chemical phenomena.

Concerning the violation of PEP for electrons, Greenberg and Mohapatra [2] in 1987 concluded that the PEP violation probability is $\leq10^{-9}$. In 1988, the Ramberg and Snow [3] experiment set a PEP violation probability for electrons $\leq1.7\times10^{-26}$. The VIP experiment [4], in 2010, improved this limit bringing it to $\leq6\times10^{-29}$.

Presently, we are considering the upgrade of the VIP setup to overcome some technical drawbacks and, for this purpose, we plan to do the following: use a much more compact system, with a higher acceptance; use a copper target capable of circulating a higher current; use Silicon Drift Detectors with a better energy resolution and with much faster time of answer, which allow the use of a veto system; reduce the background by means of better shielding.

With respect to VIP, the VIP-Upgrade geometrical gain factors are the following: higher acceptance for signal due to geometry of at least a factor 12; we can gain a factor 2 in signal by circulating a current at least of 100 A in the new cooled copper target; finally, we lose a factor 8.8/3, about 3, because of the overall target length. The combined contribution of the previous upgrades give a gain factor about 8, which enters linearly in the formula for calculating the probability of PEP violation for electrons. Besides the geometrical gain factors, we can add factors coming from background reduction (entering as square root in the overall gain for probability): a factor 4 from better energy resolution of the SDD detectors (170 eV FWHM instead of 340 eV of CCDs); a factor about 20 from the reduced active SDD surface ($6 cm^{2}$ w.r.t. $114 cm^{2}$ of VIP); a factor 5-10 from the use of better shielding and veto system; background increases with a factor about 2 due to higher efficiency of SDDs w.r.t. CCDs. Overall gain from the background reduction is about a factor 200-400, which gives a gain factor for the probability of about 15-20.

The overall gain coming from geometrical factor and reduced background is the product of the previous factors: $8\times15=120$.

In this way we expect either to find a small violation or to be able to bound the probability that PEP is violated by electrons pushing it from $\sim6\times10^{-29}$ to $\sim10^{-31}$ (this limit is obtained taking into account a data taking period similar to that of VIP, i.e. about 3-4 years).

\vfill  

\noindent{\bf References }
\begin{description}
\setlength\itemsep{-3pt}
\item{[1]} W. Pauli, Phys. Rev. {\bf 58} (1940) 716.
\item{[2]} O.W. Greenberg, R.N. Mohapatra, Phys. Rev. Lett. {\bf 59} (1987) 2507.
\item{[3]} E. Ramberg, G.A. Snow, Phys. Lett. B {\bf 238} (1990) 438.
\item{[4]} S. Bartalucci {\it et al.}, Found. Phys. {\bf 40} (2010) 765-775.
\end{description}


%% file: curceanu/curceanu.tex

%





\titl{From the Pauli Exclusion Principle violation tests (VIP experiment) to collapse models experimental investigation plans}

\name{
C.~Curceanu$^{1}$ on behalf on the VIP Collaboration
}

\adr{
$^1$ Laboratori Nazionali di Frascati dell'INFN, Via E. Fermi 40, 00044
  Frascati (Roma), Italy \\
}


A method of investigating for a possible small violation of the Pauli
Exclusion Principle (PEP) for electrons, which respects the so-called
Messiah-Greenberg superselection rule, is the search for "anomalous" X-ray
transitions in copper atoms, produced by "fresh" electrons (brought inside a
copper foil by a circulating current of 40 A) which do the Pauli-forbidden
transition to the 1s level already occupied by two electrons. 
The energy of the PEP prohibited  $2P \to 1S$  transition is shifted with respect
to the ``normal'' one by about 300 eV, giving a clear experimental signature. 
The VIP experiment is looking for the PEP violating transition with the use of
Charge Coupled Device (CCD) detectors to measure the X rays. 
In order to reduce the background, the experimental setup  was installed at
the Gran Sasso LNGS underground laboratories in 2006 and was in data taking
until 2010. 
The VIP experimental limit on the probability of PEP violation for electrons was found to be 6 $\times 10^{-29}$  [1].
Presently, the VIP collaboration is considering a major upgrade of the
experimental setup which will allow to gain other two orders of magnitude in
the probability of PEP violation limit. 
The main ideas are to use faster (than CCD) Silicon Drift Detectors combined
with a veto system, namely scintillators surrounding the setup, 
which will veto on background coming from outside the setup, and to realize a much more compact geometry, with higher acceptance.

\par We are, in parallel, considering the use of a similar experimental technique to measure the spontaneously emitted X rays, predicted  in the framework of collapse models (dynamical reduction models) [2]. Such models were put forward alternatively to the "standard" quantum mechanics' Schrodinger equation, followed by a "alla von Neumann" collapse of the wave-function, implementing a (nonrelativistic) dynamical reduction/collapse models, by modifying with a non-linear and stochastic terms  the Schrodinger equation. Baring on the importance of this conceptually new model(s), it is of utmost importance to study its experimental consequences, where the predictions are diverging from the standard equations, and to perform dedicated experiments to check it.
Today there are very few and far from complete experimental information. We aim to perform a feasibility study for a dedicated experiment to check the collapse models.

\vfill  

\noindent{\bf References }
\begin{description}
\setlength\itemsep{-3pt}
\item{[1]} C. Curceanu {\it et al.}, International Journal of Quantum Information, {\bf 9} (2011) 145.

\item{[2]} Q. Fu, Physical Review A {\bf 56} (1997) 1806.
\end{description}


%% file: dakic/dakic.tex

%





\titl{Probabilistic theories: classical, quantum and beyond quantum}

\name{
B.~Daki{\' c}$^{1}$
}

\adr{
$^1$ Faculty of Physics, University of Vienna, Boltzmanngasse 5, A-1090 Vienna, Austria \\
}


What we learn from quantum formalism is that a physical state is represented by a density operator, a mathematical object that lives in a Hilbert space. Unfortunately, it does not provide us an answer what is the ontological background of the theory, it only gives a prescription how to compute the probabilities of the measurement outcomes. It is reasonable to say that we still lack a clear operational interpretation of quantum formalism, therefore, there might be a possibility for some deeper ontology of the quantum theory. For many years there has been lot of attempts to interpret the density operator as a statistical tool that represents our lack of knowledge (of the system being measured), that is described by a more complete theory such as various hidden-variable theories~[1,2]. So far, quantum mechanics successfully survived experimental tests against such models, therefore it should have a status of genuine probabilistic theory.

In our work~[5] we employ an instrumentalist (operational) approach to study a broad class of generalized probabilistic theories where quantum mechanics is just one particular theory that one can think of. The general framework~[3,4] involves common tools that an experimentalist faces in laboratory, such as preparation, transformation and the measurement device. The main goal is to identify the set of physical principles that singles quantum theory from a vast majority of a generalized probabilistic theories.

Firstly, in analogy to quantum theory, we employ a principle of \emph{limited information content}~[6], i.e. the information capacity of systems is fundamentally limited (one-bit system contains at most one bit of information). If the finite amount of information is encoded in the state of system, then the system can reveal with certainty answers to a finite number of questions that are asked in the measurement. Therefore, the Heisenberg uncertainty and the complementarity principle come as natural consequence. The principle is valid both in quantum and classical probability theory. The whole hierarchy of ``quantum-like''theories is derived where the set of states for a two-level system is represented by an Euclidean ball in some dimension $d$, e.g. $d=1$ for a classical bit and $d=3$ for a quantum bit~[7].

To further distinguish quantum and classical theory from the full class of the probabilistic theories we adopt two additional principles: \emph{locality} and \emph{reversibility}. With the ``locality'' principle we assume that a global state of composite system can be full reconstructed from the statistics of individual systems. In other words, what determines a state of composite system is the set of marginal distributions and correlations between the subsystems.
The ``reversibility'' principle states that all the pure states can be transformed to each other by some physical, reversible transformation. It is trivial to conclude that classical theory obeys both principles. Surprisingly, this constraints leave a room for just one more theory--quantum theory itself. Finally, quantum is separated from the classical theory by the requirement that the set of reversible transformations is \emph{continuous}.

\vfill  

\noindent{\bf References }
\begin{description}
\setlength\itemsep{-3pt}
\item{[1]} J. S. Bell, Physics {\bf 1} (1964) 195--200.
\item{[2]} A. Leggett, Foundations of Physics {\bf 33} (2003) 1469--1493.
\item{[3]} L. Hardy, {\it Quantum Theory From Five Reasonable Axioms}, arXiv:quant-ph/0101012.
\item{[4]} J. Barrett, Phys. Rev. A, {\bf 75} (2007) 032304.
\item{[5]} B. Daki{\' c} and {\v C}. Brukner, in {\it ``Deep Beauty: Understanding
the Quantum World Through Mathematical Innovation''}, Eds. H. Halvorson (Cambridge University Press, 2010), pp. 365.
\item{[6]} A. Zeilinger, Foundations of Physics {\bf 29} (1999) 631--643.
\item{[7]} T. Paterek, B. Daki{\' c}, and {\v C}. Brukner, 	New J. Phys. {\bf 12}, (2010) 053037.
\end{description}


%% file: dariano/dariano.tex

%




\titl{A Quantum Digital World:\\ Quantum Fields as Quantum Automata} \name{Giacomo Mauro D'Ariano}
\adr{{\em QUIT} Group,  Dipartimento di Fisica ``A. Volta'', 27100 Pavia,  Italy, {\em http://www.qubit.it}\\
  Istituto Nazionale di Fisica Nucleare, Gruppo IV, Sezione di Pavia} Can we reduce Quantum Field
Theory (QFT) to a quantum computation? Can physics be simulated by a quantum computer? Do we believe
that a quantum field is ultimately made of a numerable set of quantum systems that are unitarily
interacting? A positive answer to these questions corresponds to substituting QFT with a theory of
quantum cellular automata (QCA): the work [1] studies this hypothesis. These investigations are part
of a large research program on {\em quantum-digitalization of physics}, with Quantum Theory as a
special theory of information [2-4],\footnote{That Quantum Theory is a theory of information has
  been shown in Ref. [3], where the theory is derived from six fundamental assumptions about how
  information is processed.}  and physics emergent from quantum-information processing [4-6].  A
QCA-based QFT has tremendous potential advantages compared to QFT, being fully {\em quantum
  ab-initio}, and free from the problems due to the continuum. Relativistic covariance at large
scale is a free-bonus of the quantum processing [6,7], with dynamics emerging from the processing.
The classical field Hamiltonian is derived from the QCA [4], opening the route to a {\em
  classicalization} procedure. The QCA reproduces the Dirac-field phenomenology at large scales, and
shows departures from QFT at the Planck-scale masses and momenta (see Fig. 1).  The
Feynman's problem (about the possibility of simulating a Fermi field with a quantum computer) is
solved with a new Jordan-Wigner transformation [1] with auxiliary qubits associated to Majorana
fields.  \def\tee{\,\mathfrak{t}\,}\def\ell{\,\mathfrak{a}\,} \def\emm{\,\mathfrak{m}\,}
\begin{figure}[h]\def\figdir{./dariano/}
\includegraphics[width=.12\textwidth]{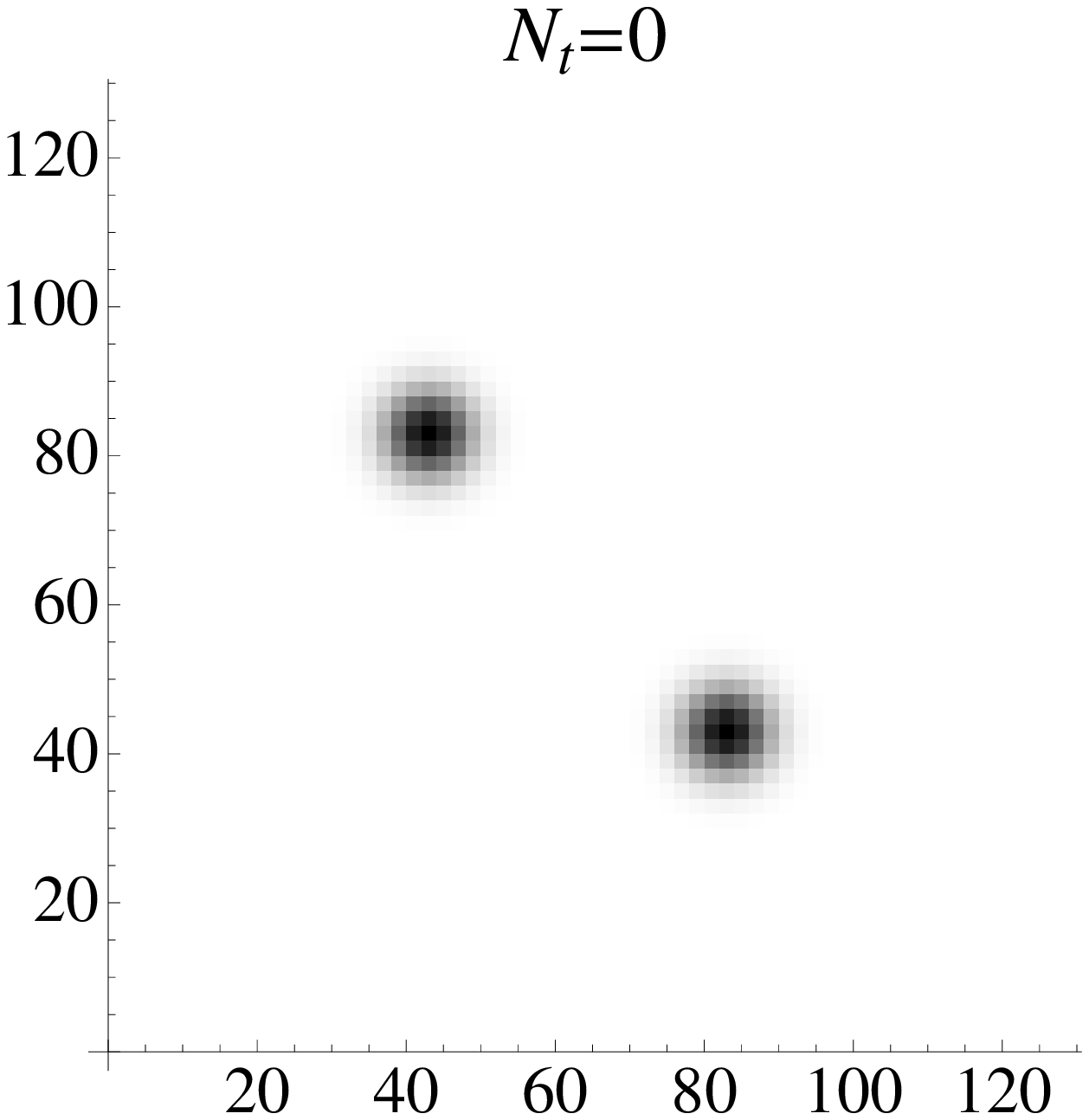} 
\includegraphics[width=.12\textwidth]{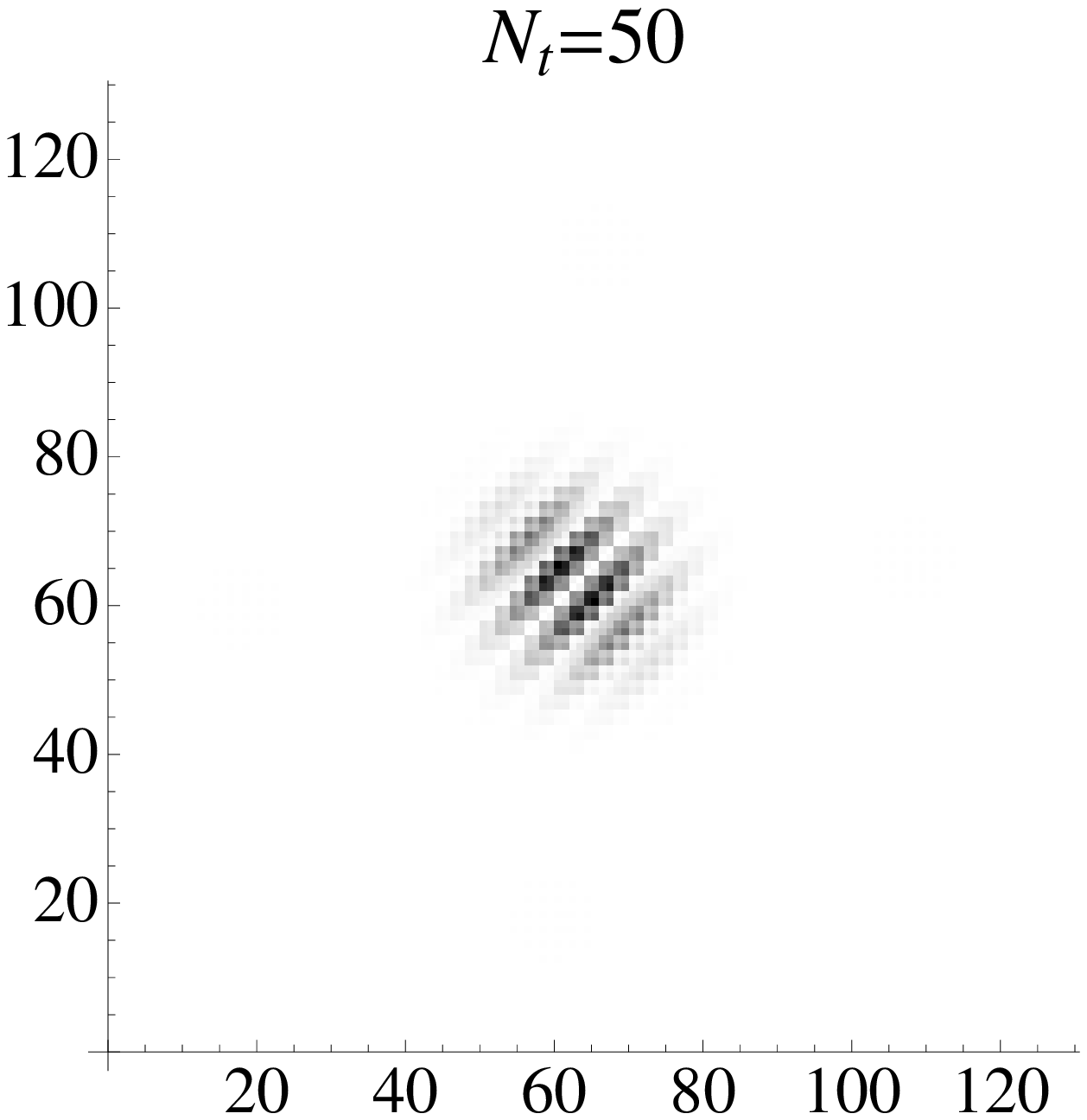} 
\includegraphics[width=.12\textwidth]{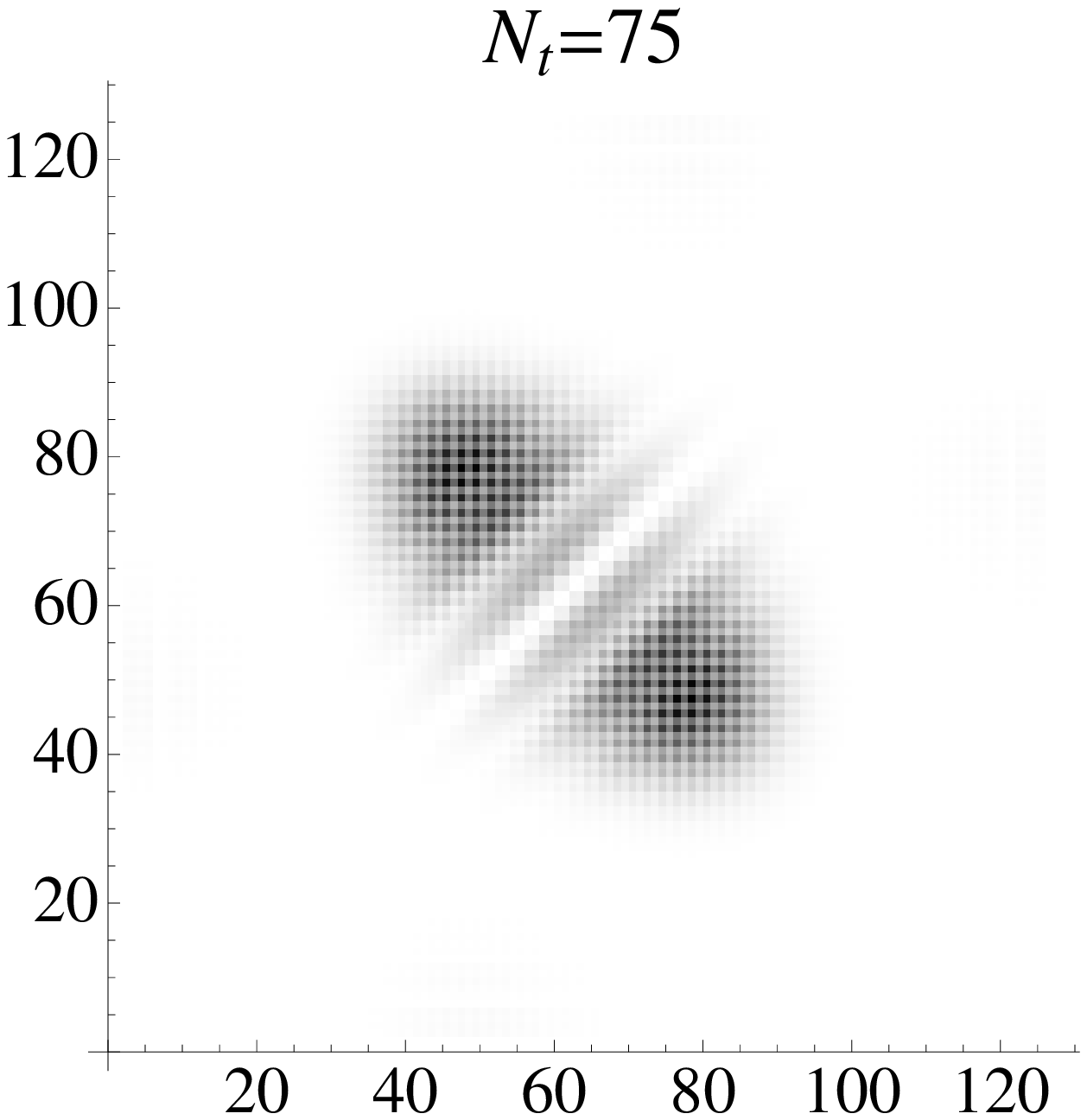} 
\includegraphics[width=.12\textwidth]{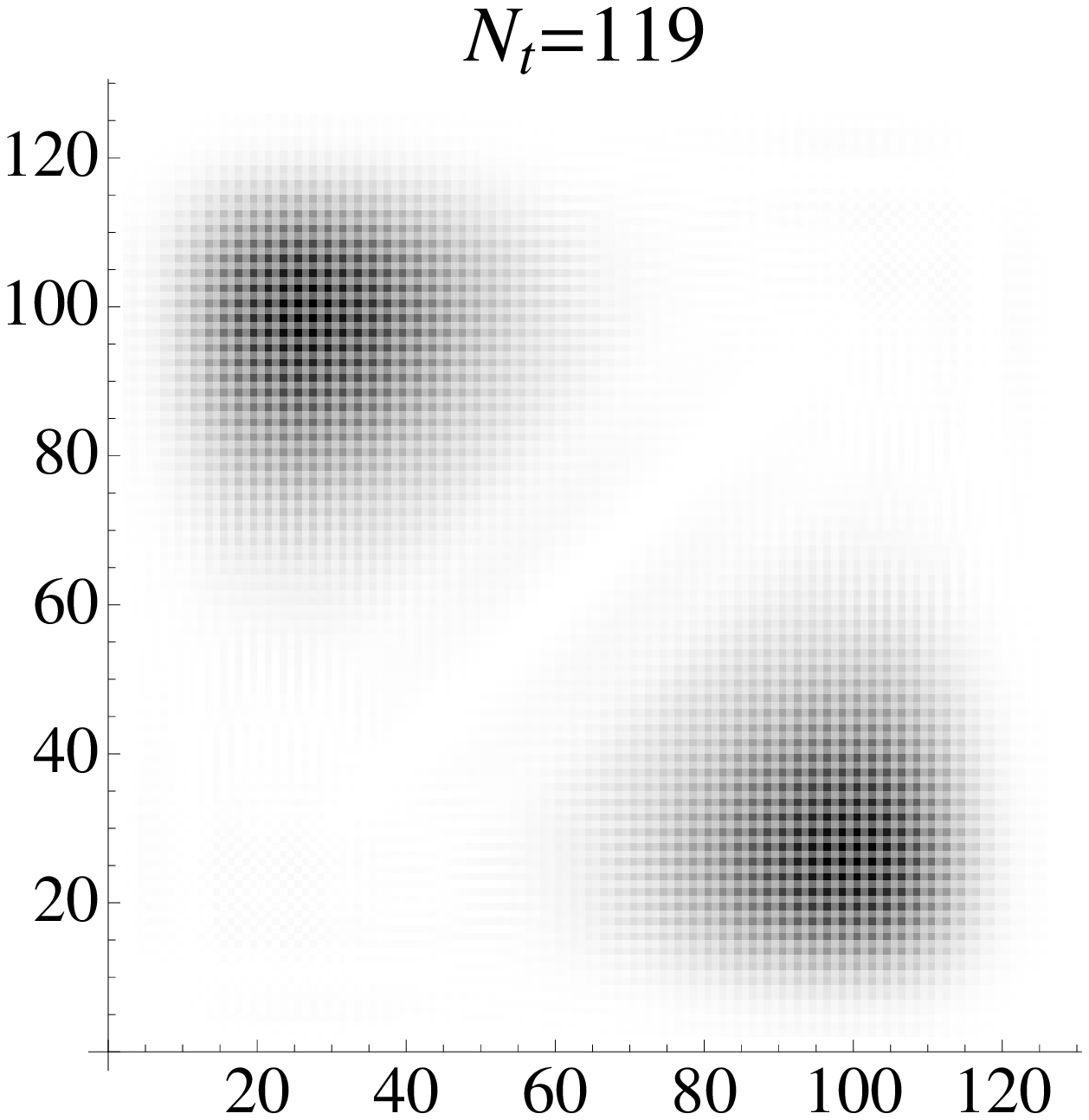} 
\includegraphics[width=.12\textwidth]{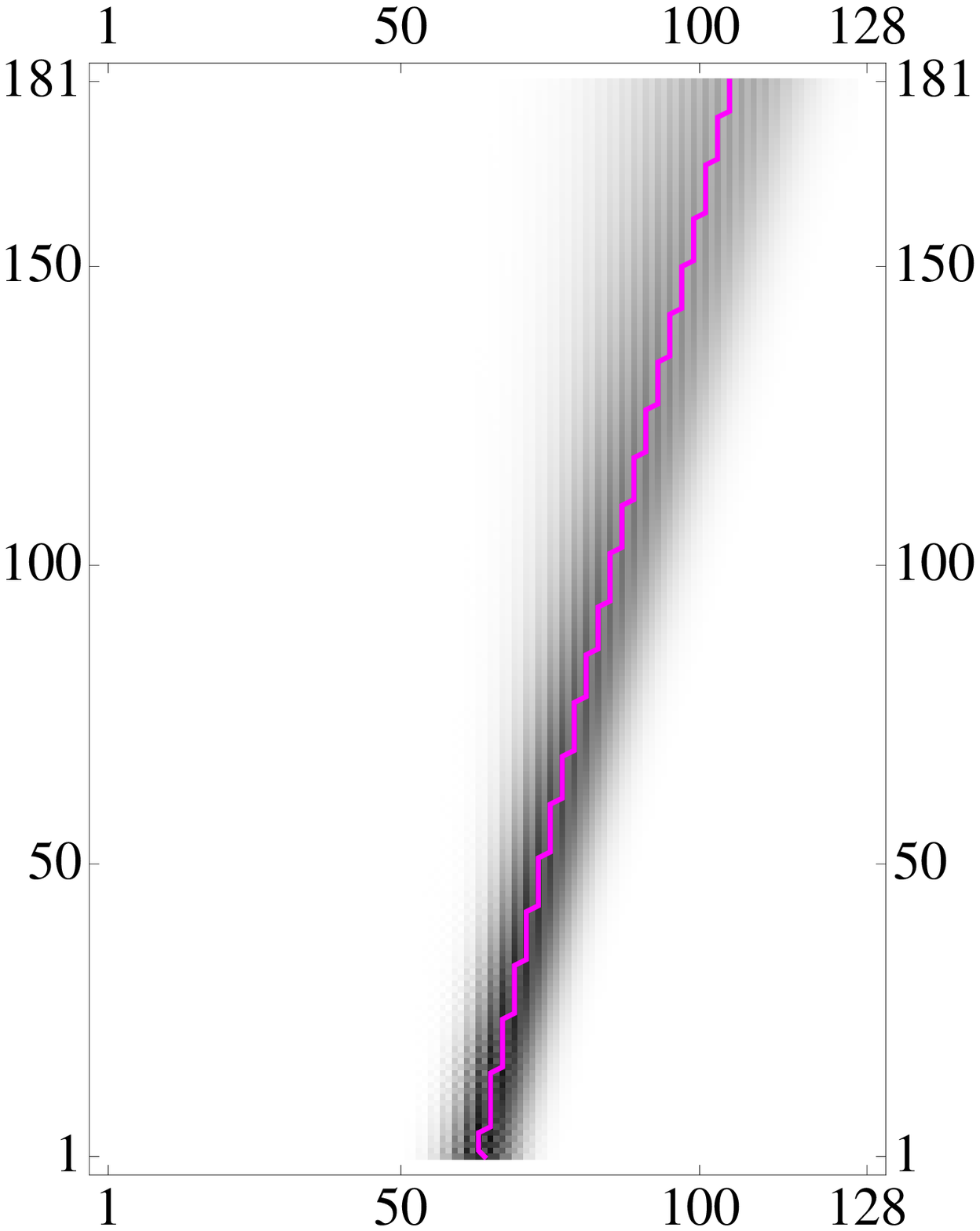} 
\includegraphics[width=.12\textwidth]{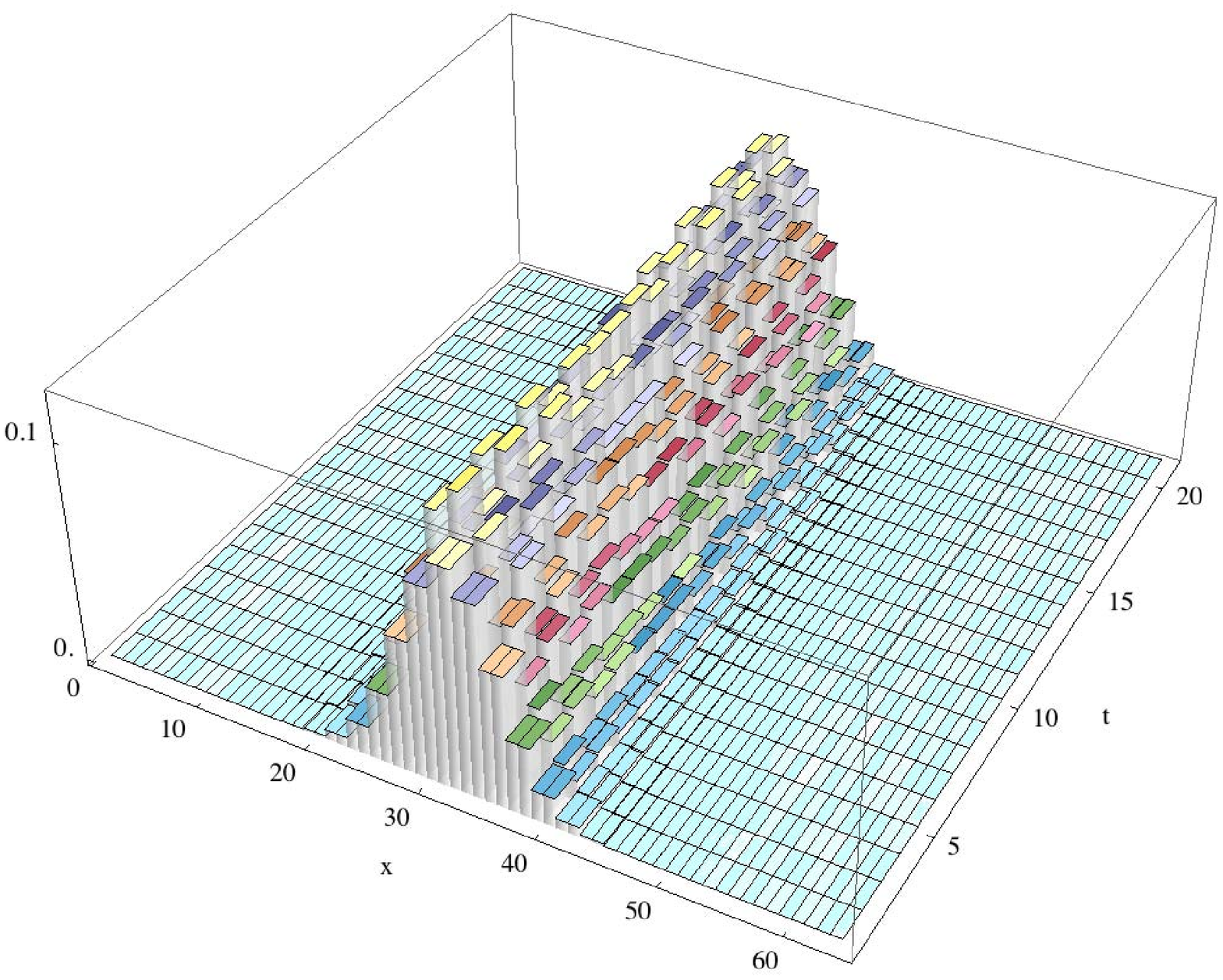} 
\includegraphics[width=.2\textwidth]{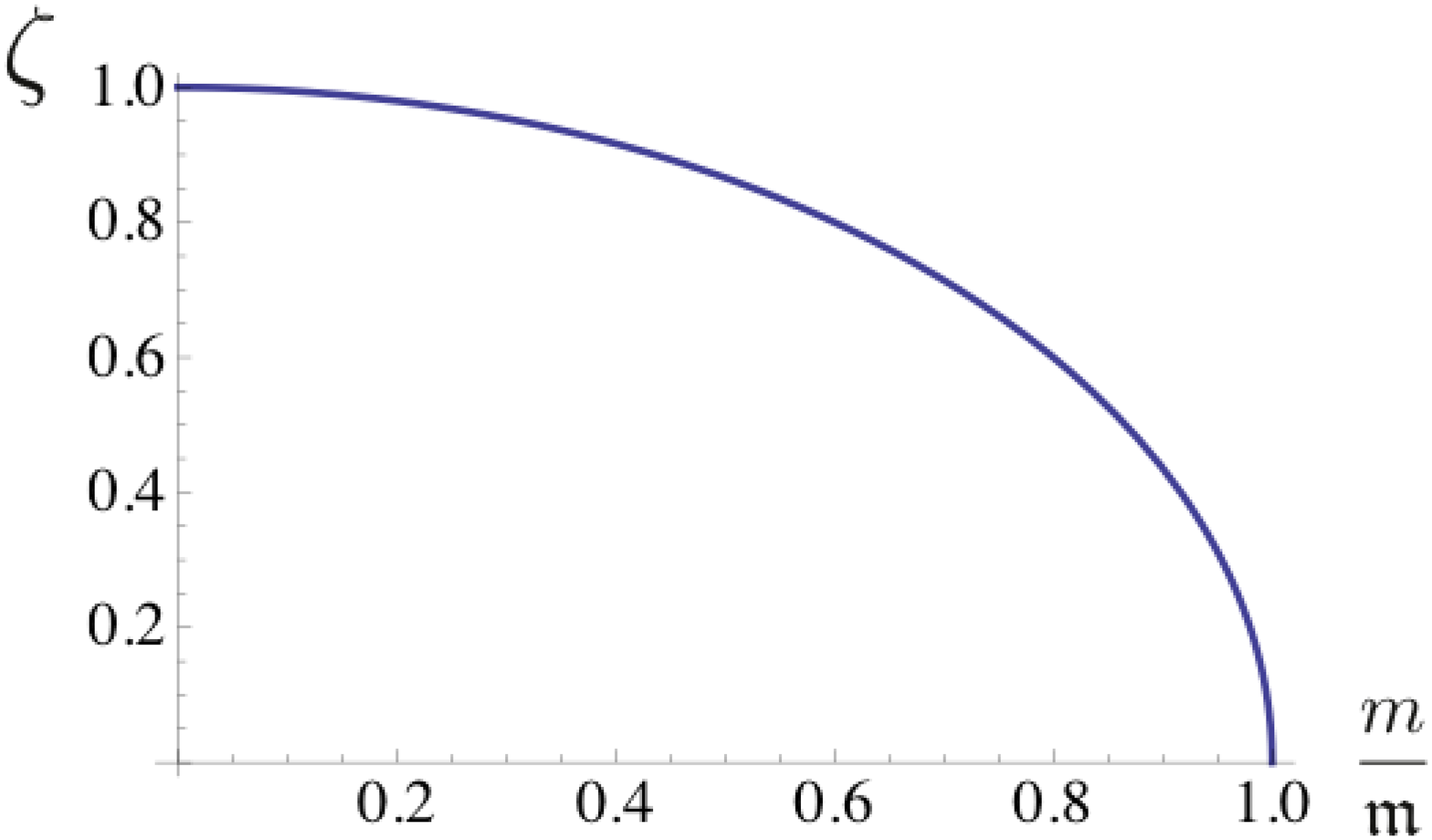}
\caption{{\footnotesize Dirac automata in 1 space dimension. {\bf Figs. 1-4:} Two-Fermion state at
 time-steps $N_t$ (Gaussian packets in collision made with 128 qubits), at
    Planckian mass and momentum ($m\simeq .92\emm$, $\emm=\hbar/(2\ell c)$ Planck mass, $\ell$
    period of the automaton). The axes are the coordinates of the two particles (symmetry along
    diagonal due to indistinguishabability).  {\bf Fig 5,6:} Details of the packet evolution.
    {\bf Fig 7:} Planckian violation of dispersion relations: inverse vacuum refraction index
    $\zeta$ vs $m/\emm$ (Ref. [4]).}}
\end{figure}

\noindent{\bf References }
\begin{description}
\setlength\itemsep{-3pt}
\item{[1]} G. M. D'Ariano, arXive:1110:6725 (2011)
\item{[2]} G. M. D'Ariano, in A.~Bokulich and G.~Jaeger eds., (Cambridge University Press, Cambridge UK, 2010)
\item{[3]} G. Chiribella, G. M. D'Ariano, P. Perinotti, Phys. Rev. A {\bf 84} 012311 (2011)
\item{[4]} G. M. D'Ariano, arXiv:1012.0756 (2010)
\item{[5]} G. M. D'Ariano, essay, FQXi Essay, Contest 2011 {\em Is Reality Digital or Analog?} {\tt
    http://fqxi.org/community/essay/winners/2011.1}
\item{[6]} M. D'Ariano, in AIP CP {\bf 1232} Quantum Theory: Reconsideration of Foundations,
 5, edited by A.~Y. Khrennikov (AIP, Melville, New York, 2010), pag 3 (also arXiv:1001.1088).
\item{[7]} G. M. D'Ariano and A. Tosini, arXiv:1109.0118
\end{description}

%% file: diaz-torres/diaz-torres.tex

%





\titl{Quantum decoherence in nuclear collision dynamics}

\name{
Alexis Diaz-Torres
}

\adr{
ECT$^{*}$, Strada delle Tabarelle 286, 38123 Villazzano (Trento), Italy \\
Email: torres@ectstar.eu}


Atomic nuclei are complex, quantum-many body systems whose internal structure is reflected in different quantum states 
associated with different types of degrees of freedom. Collective excitations (vibration and/or rotation) dominate at low energy, near the ground-state. These states are commonly employed, within a truncated model space, as a basis in coherent coupled-channels approaches to low-energy nuclear collision dynamics. However, couplings to excluded states can be essential, their effects being treated by complex potentials. Is this a complete description of open, quantum nuclear dynamics? Does it include effects of quantum decoherence? Can decoherence effects be manifested in reaction observables?

I reported on innovative studies on decoherence in nuclear collisions [3-5], which are largely motivated by systematic disagreements between high-precision measurements of fusion [1] and scattering [2] of nuclei and theoretical predictions based on the coherent quantal description such as the standard coupled channels framework.

These studies [3-5] exploit the time propagation of a coupled-channels density matrix for the collective degrees of freedom (reduced system), including the center-of-mass motion of the interacting nuclei. The density matrix obeys the Lindblad master equation, where physically motivated operators describe the irreversible couplings of the reduced system with the high-density bath of single-particle excitations (environment). This bath emerges in the course of a low-energy collision, when the nuclei are close to each other, and nuclear and Coulomb interaction fields are strong. The approach accounts for decoherence and dissipation, going beyond the usually employed complex potential model where decoherence is not included [5]. Model calculations [4] indicate that decoherence effects may be seen in heavy-ion scattering through the angular distribution of inelastic excitations of the colliding nuclei. Decoherence affects the quantum interference among the different partial waves, changing by a few degrees the minima of these angular distributions at backward scattering angles.  

This innovative approach has a wide range of applications. In the nuclear physics context, it can be applied to understand both quantum tunnelling of exotic nuclei and the formation of superheavy elements, as well as to investigate quantum nuclear dynamics driven by super-intense and ultra-short laser pulses. These are relevant for experimental nuclear physics programmes at new facilities such as SPIRAL2 and ELI. The approach also finds applications in other areas, for instance, in understanding light-harvesting and transport of energy in photobiological systems.

\vfill  

\noindent{\bf References }
\begin{description}
\setlength\itemsep{-3pt}
\item {[1]} M. Dasgupta {\it et al.}, Phys. Rev. Lett. {\bf 99} (2007) 192701.
\item {[2]} M. Evers {\it et al.}, Phys. Rev. C {\bf 78} (2008) 034614.
\item{[3]} A. Diaz-Torres {\it et al.}, Phys. Rev. C. {\bf 78} (2008) 064604.
\item{[4]} A. Diaz-Torres, Phys. Rev. C. {\bf 82} (2010) 054617.
\item{[5]} A. Diaz-Torres, Phys. Rev. C {\bf 81} (2010) 041603(R).
\end{description}


%% file: didomenico/didomenico.tex

%





\titl{$CPT$ symmetry, Quantum Mechanics, and Neutral kaons}

\name{
A.~Di~Domenico$^{1,2}$
}

\adr{
$^1$ Dipartimento di Fisica, Sapienza Universit\`a di Roma, Rome, Italy \\
$^2$ INFN Sezione di Roma, Rome, Italy \\
}

The neutral kaon system offers a unique possibility to perform fundamental tests 
of the basic principles of quantum mechanics and
of $CPT$ symmetry [1].
The most recent limits obtained by the KLOE experiment at the DA$\Phi$NE
$e^+e^-$ collider on several kinds
of possible decoherence and $CPT$ violation 
mechanisms [2,3], which in some cases might be justified in a quantum
gravity framework,
show
no
deviation from the expectations of quantum mechanics and $CPT$ symmetry, while the precision of the measurements, in some cases, 
reaches the interesting 
Planck scale region. 
Prospects for this kind of experimental studies
at KLOE-2 foresee an improvement in the precision of about 
one order of magnitude [4].

\vfill  

\noindent{\bf References }
\begin{description}
\setlength\itemsep{-3pt}
\item{[1]} {\it Handbook on neutral kaon interferometry at a $\phi$-factory}, A.~Di~Domenico ed., Frascati Physics Series {\bf 43}, INFN-LNF, Frascati, 2007.
\item{[2]} F.~Ambrosino et al., KLOE collaboration,  Phys. Lett. B {\bf 642} (2006) 315.
\item{[3]} A.~Di Domenico and the KLOE collaboration, Journal of Physics: Conf. Series {\bf 171} (2009) 012008.
\item{[4]} G.~Amelino-Camelia et al., Eur. Phys. J. C {\bf 68} (2010) 619.
\end{description}


%% file: ferialdi/ferialdi.tex

%





\titl{Non-Markovian features in quantum dynamics}

\name{
L.~Ferialdi$^{1}$
}

\adr{
$^1$ Department of Physics, University of Trieste, 34151 Trieste, Italy}


Open quantum systems are those systems which interact with surrounding environment. These systems are typically described by means of Markovian dynamics. Such dynamics are characterized by a well defined structure, of the Lindblad type, whose mathematical properties are known. Markovian dynamics are analyzable with such an accuracy because they lack of memory effects: the dynamics does not depend on the past history of the system.

 However, there are many physical situations in which the Markovian description is not sufficient anymore, among which ultra-fast processes in chemistry [1], energy transfer and light harvesting in the photosynthesis process [2].

 The difficulty in treating non Markovian dynamics lies in the fact that these, unlike Markovian ones, contain memory terms about the past history of the system. Many important steps forward have been taken and several phenomenological models have been proposed~[3,4,5].
 
 Among various approaches proposed to study non Markovian dynamics, stochastic differential equations (SDEs) represent a particularly interesting method. In this approach the environment is described by means of suitable stochastic terms (or noise terms) added to the usual Schr\"odinger evolution. In particular, for non Markovian dynamics the interaction with the system is modeled with noises that have a general time correlation function~[3,6].
 
 SDEs are also used in the description of collapse models. These models were introduced to solve the measurement problem in quantum mechanics. Basically, the idea is that the wave function undergoes a spontaneous collapse which is random both in space and time. In order to include these effects in the standard evolution, one has to modify the Schr\"odinger equation introducing some non-linear and stochastic terms, obtaining a SDE.

The non-Markovian QMUPL model represents an interesting model, being physically meaningful and mathematically analyzable in detail. Such model is described by the following SDE:
 \begin{equation}
 \frac{d}{dt}\phi_t=\left[-\frac{i}{\hbar}H+\sqrt{\lambda}q w(t)-2\sqrt{\lambda}q\int_0^t ds D(t,s)\frac{\delta}{\delta w(s)}\right]\phi_t\,.
 \end{equation}
This equation have been solved using the path-integral formalism, and the features of the dynamics have been analyzed in detail [3,7]. This shows that the SDEs are a powerful mathematical approach to study non-Markovian dynamics, and that they allow for a great insight in understanding these dynamics.

\vfill  

\noindent{\bf References }
\begin{description}
\setlength\itemsep{-3pt}
\item{[1]} E. Gindensperger, I. Burghardt, L. S. Cederbaum,
J. Chem. Phys. {\bf 124}, 144103 (2006).
\item{[2]} M. Thorwart {\it et al.}, Chem. Phys. Lett. {\bf 478}, 234 (2009).
\item{[3]}  A. Bassi, L. Ferialdi, Phys. Rev. Lett. {\bf 103}, 050403 (2009).
\item{[4]} J. Piilo {\it et al.}, Phys. Rev. Lett. {\bf 100}, 180402 (2008).
\item{[5]} K. H. Hughes, C. D. Christ, I. Burghardt, J. Chem. Phys. {\bf 131}, 024109 (2009).
\item{[6]} L.~Di{\'o}si, W.~T. Strunz, Phys. Lett. A \textbf{235}, 569 (1997).
\item{[7]} A. Bassi, L. Ferialdi, Phys. Rev. A {\bf 80}, 012116 (2009).
\end{description}


%% file: floreanini/floreanini.tex

%





\titl{Quantum contextuality and identical particles}

\name{
F. Benatti$^{1,2}$, R. Floreanini$^2$, M. Genovese$^3$, S. Olivares$^1$
}

\adr{
$^1$ Dipartimento di Fisica, Universit\`a degli Studi di
Trieste, I-34151 Trieste, Italy\\
$^2$ Istituto Nazionale di Fisica Nucleare, Sezione di
  Trieste, I-34151 Trieste, Italy \\
$^3$ INRIM, Strada delle Cacce 91, I-10135 Torino, Italy
}


\bigskip 
A physical theory is called noncontextual if the measurement of a given observable
does not depend on whether other commuting observables 
are simultaneously measured, {\it i.e.} it does not depend on the context. 
Unlike classical physics,
quantum mechanics turns out to be a contextual theory [1-3]. Various experiments have been
performed in order to directly test this property, mainly using single spin-1/2 particles or photons.
Unlike Bell-locality tests based on entangled physical systems, the above experiments 
involve two degrees of freedom of the same physical system: typically, one degree of 
freedom is translational, while the other is internal, spin or helicity.
Most of the tests check whether inequalities of Clauser-Horne-Shimony-Holt (CHSH)
type, applied now to local, single-system settings, are violated or not.

\medskip

In systems made of $N$ identical bosonic particles, tests of quantum contextuality 
can not be exhibited using single-particle observables, as adopted in the
approaches so far considered in the literature: collective observables built with
suitable multiboson operators are needed instead [4].
More specifically, using maximally entangled NOON-like
states, one can show that inequalities of CHSH type
can not be violated if constructed with standard
single-boson observables; this is a peculiarity of such systems and can
be ultimately traced back to the indistinguishability of the constituents.  
In order to make quantum contextuality apparent in such systems, 
one has to use instead
collective observables built with suitable multiboson operators [5].
Indeed, with such observables one can construct CHSH like inequalities
that are maximally violated by the same NOON-like states [4].

\medskip

A physical implementation based on an interferometric scheme able 
to experimentally test these results can be realized within
quantum optics. In this context, the two degrees of freedom necessary
to build the proper observables entering the CHSH inequality
can be taken to be the photon path (momentum) and polarization.
However, note that operations acting on the $N$-photon states 
as a whole are needed; as a consequence, linear optical passive 
devices cannot directly be used. In spite of this, an implementation
based on a Mach-Zehnder-like interferometric
scheme can be easily conceived [4]; in such a scheme,
the usual beam splitters are replaced by the ``quantum
beam splitters'', where a nonlinear medium is inserted 
in one of the two arms. These results will surely open
new perspectives in the study of quantum contextuality in
mesoscopic systems and stimulate new insights
for further experimental tests.

\vfill  

\noindent{\bf References }
\begin{description}
\setlength\itemsep{-3pt}
\item{[1]} N.D. Mermin, Rev. Mod. Phys. \textbf{65} (1993) 803 
\item{[2]} M. Genovese, Phys. Rep. \textbf{413} (2005) 319 
\item{[3]} F. Benatti {\it et al.} J. Phys. A: Math. Gen. {\bf 35} (2002) 4955 
\item{[4]} F. Benatti {\it et al.}, Phys. Rev. A {\bf 84} (2011) 034102 
\item{[5]} M. Rasetti, Int. J. Theor. Phys. {\bf 5} (1972) 377 
\end{description}


%% file: fomenko/fomenko.tex

%





\titl{Study of rare processes with the Borexino detector}

\name{
A. Derbin$^{1}$, K.~Fomenko$^{2}$ \\
on behalf of Borexino Collaboration: \\
\ \\
{\small
G.~Bellini, J.~Benziger, D.~Bick, S.~Bonetti, M.~Buizza Avanzini, B.~Caccianiga, L.~Cadonati,
F.~Calaprice, C.~Carraro, P.~Cavalcante, A.~Chavarria, D.~D’Angelo, S.~Davini, A.~Derbin,
A.~Etenko, F.~von~Feilitzsch, K.~Fomenko, D.~Franco, C.~Galbiati, S.~Gazzana, C.~Ghiano,
M.~Giammarchi, M.~G¨oger-Neff, A.~Goretti, L.~Grandi, E.~Guardincerri, S.~Hardy, Al.~Ianni,
An.~Ianni, V.~Kobychev, D.~Korablev, G.~Korga, Y.~Koshio, D.~Kryn, T.~Lewke, E.~Litvinovich,
B.~Loer, P.~Lombardi, I.~Machulin, S.~Manecki, W.~Maneschg, G.~Manuzio, Q.~Meindl,
E.~Meroni, L.~Miramonti, M.~Misiaszek, D.~Montanari, P.~Mosteiro, V.~Muratova, L.~Oberauer,
M.~Obolensky, F.~Ortica, M.~Pallavicini, L.~Papp, L.~Perasso, S.~Perasso, A.~Pocar,
R.S.~Raghavan, G.~Ranucci, A.~Razeto, A.~Re, A.~Romani, A.~Sabelnikov, R.~Saldanha, C.~Salvo,
S.~Sch¨onert, H.~Simgen, M.~Skorokhvatov, O.~Smirnov, A.~Sotnikov, S.~Sukhotin, Y.~Suvorov,
R.~Tartaglia, G.~Testera, D.~Vignaud, R.B.~Vogelaar, J.~Winter, M.~Wojcik, A.~Wright,
M.~Wurm, J.~Xu, O.~Zaimidoroga, S.~Zavatarelli, and G.~Zuzel
}
}

\adr{
$^1$ Neutron Research Department, St. Petersburg Nuclear Physics Institute RAS, 188300 Gatchina, Russia \\
$^2$ Laboratory for Nuclear Problems, Joint Institute for Nuclear Research, 141980 Dubna, Russia \\
}


The status of the search for rare processes with the Borexino detector and it's prototype (CTF)
is reviewed. We present results of search for electron decay $e \rightarrow \gamma + \nu$~[1], 
violation of the Pauli exclusion principle in nuclei, 5.5 MeV axions from $p + d \rightarrow {\rm{^3}He} + A$ 
reaction on the Sun and magnetic moments of solar neutrinos. New limits on the probability of non-Paulian 
transitions in $\rm^{12}C$ nuclei [2], on the axion-electron, axion-photon and axion-nucleon coupling constants [3]
and on the neutrino magnetic moments [4] are presented.

\vfill  

\noindent{\bf References }
\begin{description}
\setlength\itemsep{-3pt}
\item{[1]} H.O.Back {\it et al.} Borexino Collaboration, 
                          { ``Search for electron decay mode  $e \rightarrow \gamma = \nu$ with the prototype of 
                                 Borexino detector"}, {\it Phys.Lett.}~{\bf B} 525, 29 (2002).

\item{[2]} G.Bellini {\it et al.} Borexino Collaboration, 
                          { ``New experimental limits on the Pauli forbidden transitions in $\rm^{12}C$ 
                                 nuclei obtained with 485 days Borexino data"}, {\it Phys.Rev.}~{\bf C} 81, 034317 (2010).

\item{[3]} G.Bellini {\it et al.} Borexino Collaboration, 
                          { ``Search for solar axions produced in $p(d,{\rm ^3He})A$ reaction with the Borexino Detector"}, 
                          to be submitted to {\it Phys.Rev.}~{\bf D}.

\item{[4]} C.Arpesella {\it et al.} Borexino Collaboration, 
                             { ``Direct Measurement of the $\rm^7Be$ Solar Neutrino Flux with 192 Days of Borexino Data"},
                             {\it Phys.Rev.Lett.}~{\bf 101}, 9 (2008).
\end{description}


%% file: gerlich/gerlich.tex

%





\titl{Matter-wave interferometry and metrology with macromolecules}

\name{
S.~Gerlich$^{1}$, S.~Eibenberger$^{1}$, J.~T\"uxen$^{2}$, M.~Mayor$^{2,3}$, M.~Arndt$^{1}$
}

\adr{
$^1$ University of Vienna, Vienna Center for Quantum Science and Technology
(VCQ), Faculty of Physics, Boltzmanngasse 5, 1090 Vienna, Austria \\
$^2$ Department of Chemistry, University of Basel, St Johannsring 19,
CH-4056 Basel, Switzerland \\
$^3$ Karlsruhe Institute of Technology, Institute for Nanotechnology,
PO Box 3640, 76021 Karlsruhe, Germany
}


Matter-wave interferometry with molecules of increasing size, mass, and complexity allows the exploration of the frontiers of quantum mechanics. We employ a Kapitza-Dirac-Talbot-Lau interferometer (KDTLI) [1,2] to investigate the quantum wave nature of organic molecules in an unprecedented mass range. In our experiments we have so far been able to demonstrate the quantum delocalization of molecules composed of up to 430 atoms, with a mass of nearly 7000 atomic mass units, a maximal size of up to 60\AA, and de Broglie wavelengths down to one picometer [3].

A KDTLI consists of a set of three consecutive gratings and operates in the near field. While silicon nitride masks are employed as the first and the third grating, the central grating is implemented as an optical grating, which helps to avoid the highly dispersive van-der-Waals interaction between the grating walls and the studied molecules. This, together with its very advantageous scaling behavior with respect to the de Broglie wavelength, makes a KDTLI ideally suited for the thermal beams and broad velocity distributions of effusive sources. In addition, a KDTLI also accepts incoherent particle beams, which eliminates the need for collimators and thus increases the transmitted particle flux by several orders of magnitude over far-field experiments. Given the generally low brightness of the sources and the low efficiency of the detectors for massive molecules, this is an essential advantage over conventional interferometer types.

A KDTLI is highly sensitive to external forces and can be used to perform precision measurements of diverse molecular properties such as the electric polarizability, the susceptibility, or the electric dipole moment [4]. We present, how the measurement of these parameters can be used to enhance mass spectrometry [5] or to identify different isomers [6]. Matter-wave interferometry also gives insight into internal molecular dynamics, such as conformational changes [7]. The information about the internal states can be exclusively obained through conservative interactions and thus allows the full preservation of the quantum delocalization in position space.

\vfill  

\noindent{\bf References }
\begin{description}
\setlength\itemsep{-3pt}
\item{[1]} S. Gerlich {\it et al.}, Nature Physics {\bf 3} (2007) 711.
\item{[2]} K. Hornberger {\it et al.}, New J. Phys. {\bf 11} (2009) 043032.
\item{[3]} S. Gerlich {\it et al.}, Nature Commun. {\bf 2} (2011) 263.
\item{[4]} S. Eibenberger {\it et al.}, New J. Phys. {\bf 13} (2011) 043033.
\item{[5]} S. Gerlich {\it et al.}, Angew. Chem. Int. Ed. {\bf 47} (2008) 6195.
\item{[6]} J. T\"uxen {\it et al.}, Chem. Comm. {\bf 46} (2010) 4145.
\item{[7]} M. Gring {\it et al.}, PRA {\bf 81} (2010) 031604(R).

\end{description}


%% file: gouanere/gouanere.tex

%





\titl{A search for the de Broglie particle internal clock by means of electron channeling}

\name{
Michel Gouanere$^{1}$
}

\adr{
$^1$ IPNL-IN2P3-CNRS: UMR5822, Universit\'{e} Claude Bernard, Lyon I, France 

}


The particle internal clock conjectured by de Broglie in 1924 was investigated 
in a channeling experiment using a beam of $\sim 80$ MeV electrons aligned 
along the $<110>$ direction of a 1 $\mu$m thick silicon crystal. 
Some of the electrons undergo a rosette motion, in which they interact with a single atomic row. 
When the electron energy is finely varied, the rate of electron transmission 
at 0{\kern-.2em\r{}\kern-.3em} shows a 8\% dip within 0.5\% of the resonance energy, 
80.874 MeV, for which the frequency of atomic collisions matches the electron's internal 
clock frequency. An early experiment in crystals to detect this very high frequency 
(ALS Saclay 1980) has been published with positive result [1].  
An experiment is underway at Fracasti (BTF) to confirm it [2].

\vfill  

\noindent{\bf References }
\begin{description}
\setlength\itemsep{-3pt}
\item{[1]} P. Catillon, N. Cue, M.J. Gaillard, R. Genre, M. Gouan\`{e}re, R. G.Kirsch, J.C. Poizat, 
J. Remillieux, L. Roussel and M. Spighel. Found. Physics {\bf 38} (2008) 659.

\item{[2]} C. Ray, M. Bajard, R. Chehab, M. Chevallier, C. Curceanu, S. Dabagov, 
D. Dauvergne, M.Gouan\`{e}re, R. Kirsch, J.C. Poizat, J. Remillieux , and E. Testa,
{\it RICCE2 experiment: Research of Internal Clock by Channeling of Electrons},
\url{http://www.lnf.infn.it/acceleratori/public/BTF_user/RICCE/RICCE.pdf} 
\end{description}


%% file: hayrapetyan/hayrapetyan.tex

%





\titl{Probability amplitudes of two-levels atoms beyond the dipole
       approximation}

\name{
A.~G.~Hayrapetyan$^{1,2}$, S.~Fritzsche$^{3,4}$ }

\adr{ $^1$ Physikalisches Institut, Universit\"a{}t Heidelberg,
             D-69120 Heidelberg, Germany \\
$^2$ Max-Planck-Institut f\"ur Kernphysik, Postfach 103980,
             D-69029 Heidelberg, Germany \\
$^3$ Department of Physics, P.O.\ Box 3000,
             Fin-90014 University of Oulu, Finland \\
$^4$ GSI Helmholtzzentrum f\"ur Schwerionenforschung,
            D-64291 Darmstadt, Germany
}

In scientific literature, atom-field interaction mainly is
considered within the dipole \textit{and} rotating wave
approximations. The dipole approximation (DA) is applicable for a
wide range of electromagnetic radiation spectrum, from
radio-frequencies (still) to ultraviolet ones. Whereas, within an
increase of an atomic number the energy gap between two atomic levels
becomes larger. Hence, the frequency of emitted (absorbed) photon
increases and obtains values in ultraviolet and even in soft X-ray
region. In this case, atoms can not be considered as point-like,
their wave functions are smeared over some region whose radius still
is less but comparable to the wavelength of the field. Therefore,
the necessity of examination of an atom-field interaction beyond DA
arises.

The (resonant) interaction of an atom with the radiation field leads
to Rabi oscillations of the population which depend both on atomic
number and the polarization properties of the field. Beyond DA, we
show how an increase of atomic number leads to the suppression of
the population inversion. Moreover, application of a circularly
polarized field yields stronger Rabi oscillations whose frequency is
exactly two times larger than what the Rabi frequency would be for a
linearly polarized field.

\begin{figure}[h]
\centering
  \includegraphics[width=16cm,height=6cm]{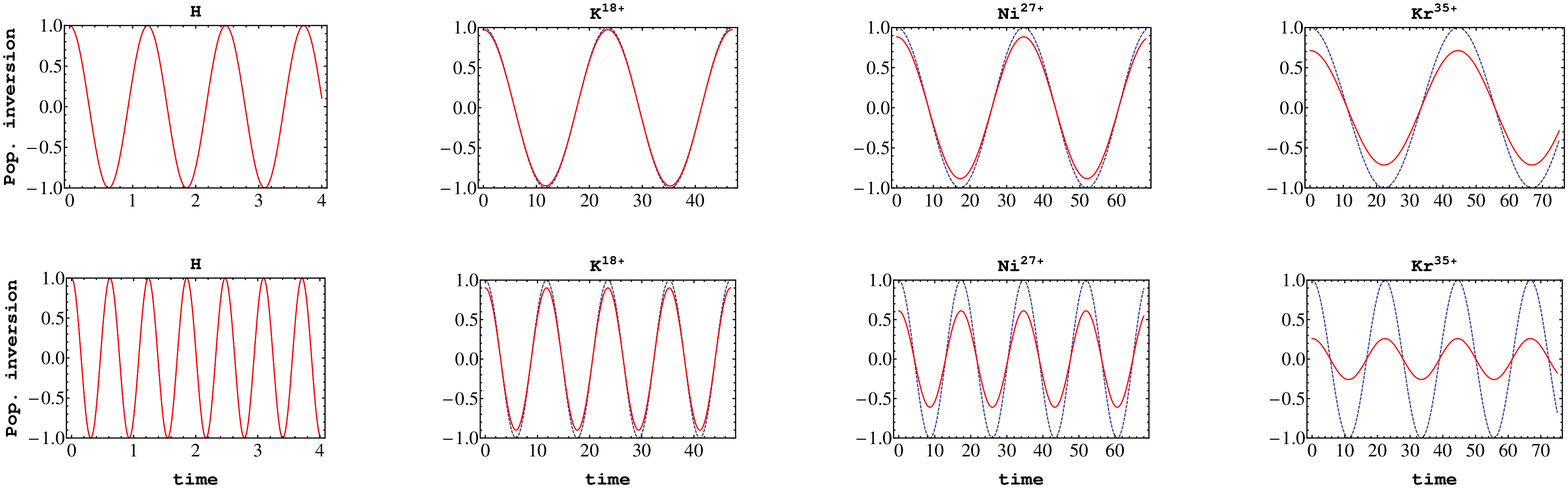}
  \caption{Population inversion for
Hydrogen, hydrogen-like Potassium, Nickel and Krypton. Upper plots
correspond to the interaction of atomic states with linearly
polarized radiation field, and the lower plots to the interaction of
an atom with left-circularly polarized radiation field ($\epsilon =
-1$). The dashed-blue line corresponds to the population inversion
in DA, and the solid-red line stands for the population inversion
beyond DA. The time is in units of $10^{-15}$ s.}
\end{figure}

We succeeded in solving analytically the Schr\"odinger equation in
the rotating wave approximation both in the absence and presence of
damping of atomic states. Moreover, the obtained solutions can be
constructed also beyond the \textit{rotating wave approximation}. In
fact, this can be a good basis for further research and
consideration of more precise phenomena.

\vfill  


%% file: hiesmayr/hiesmayr.tex

%





\titl{Bell's [Un]Speakables in the Neutral Kaon System}

\name{
B.C.~Hiesmayr
}

\adr{
University of Vienna, Faculty of Physics, Boltzmanngasse 5, 1090 Vienna, Austria\\
}


John Stuart Bell was one of the leading expositors and interpreters of modern quantum theory. He is particularly famous for his discovery of the crucial difference between the predictions of conventional quantum mechanics and the implications of locality and realism, nowadays known as Bell inequalities. The purpose of this short note is to show that systems in high energy physics can also be considered to discuss basic questions of quantum mechanics and, moreover, that certain features of entanglement only reveal themselves in such system. Certainly, to obtain the full picture of entanglement and its numerous manifestation one has to study also systems at different energy scales!

Indeed, also for strange mesons entangled states
can be obtained, in analogy to the entangled spin up and
down pairs, or $H$ and $V$ polarized photon pairs. Such
states are for example produced at the DAPHNE machine in Italy. There, a kaon $K^0$-antikaon $\bar K^0$ pair is described at
the time $t=0$ by the entangled antisymmetric Bell state,
$$ |\psi^-\rangle=\frac{1}{\sqrt{2}}\lbrace | K^0\rangle\otimes|\bar K^0\rangle-|\bar K^0\rangle\otimes|K^0\rangle\rbrace\;.$$ To describe the time evolution of this entangled state is involved, subtle and can be handled with different theoretical frameworks [3-6].
In Ref.~[1] a method is developed how the entanglement of such systems in high energy at accelerator facilities can be tested. It was used for the analyzes of the data of the KLOE detector at the DAPHNE machine [2]. The results demonstrate that the purity of the state is given to a high degree. Revealing the nonlocality for these systems is involved. In Ref.~[1] the authors
show that set of Bell inequalities lead to $\delta=0$,
where $\delta$ is the famous $\mathcal{CP}$ violating parameter in mixing ($\mathcal{C}$\dots charge conjugation, $P$\dots parity). Experimentally,
the value $\delta\approx 10^{-3}$ is in clear contradiction to the value
required by the set of Bell inequalities, i.e. by the premises of local realistic
theories. Herewith, two different concepts, nonlocality and symmetries in Particle Physics, get in a puzzling relation, clearly only available for this system. Unfortunately, this set of Bell inequalities cannot be directly experimentally verified due to experimental limitations.

\textbf{A conclusive direct experimental test of the nonlocal features of these systems entangled in strangeness is still missing and evidently needed! Its difficulty stems from the fact that in such experiments one has only access to a restricted class of measurable observables and that these systems oscillating in time are also decaying.}


\vfill  

\noindent{\bf References }
\begin{description}
\setlength\itemsep{-3pt}
\item{[1]} R.A. Bertlmann, W. Grimus, B.C. Hiesmayr, Phys. Rev. D {\bf 60} (1999) 114032.
\item{[2]} F. Ambrosino et al., KLOE collaboration, Phys. Lett. B {\bf 642} (2006) 315 ; Antonio Di Domenico, J. Phys.: Conf. Ser. {\bf 171} (2008) 1.
\item{[3]} R.A. Bertlmann, W. Grimus, B.C. Hiesmayr, Phys. Rev. A {\bf 73} (2006) 054101.
\item{[3]} R.A. Bertlmann, B.C. Hiesmayr, Phys. Rev. A {\bf 63} (2001) 062112.
\item{[4]} B.C. Hiesmayr, European Physical Journal C {\bf 50} (2007) 73.
\item{[5]} R.A. Bertlmann, A. Bramon, G. Garbarino, B.C. Hiesmayr, Phys. Lett. A {\bf 332} (2004) 355.
\item{[6]} B.C. Hiesmayr, Found. of Phys. Lett. {\bf 14} (2001) 231 .

\end{description}


%% file: liang/liang.tex

%





\titl{ Finite-speed causal-influence models \\for quantum theory lead to superluminal signaling}

\name{
Jean-Daniel Bancal$^{1}$, Stefano Pironio$^{2}$, Antonio Ac\'in$^{3,4}$,  \\Yeong-Cherng Liang$^{1}$, Valerio Scarani$^{5,6}$, and Nicolas Gisin$^{1}$
}

\adr{
$^1$ Group of Applied Physics, University of Geneva, Switzerland\\
$^2$ Laboratoire d'Information Quantique, Universit\'e Libre de Bruxelles, Belgium\\
$^3$ CFO-Institut de Ci\`encies Fot\`oniques, Av. Carl Friedrich Gauss 3, E-08860 Castelldefels (Barcelona), Spain\\
$^4$ ICREA-Instituci\'o Catalana de Recerca i Estudis Avan{\c c}ats, Lluis Companys 23, E-08010 Barcelona, Spain\\
$^5$ Centre for Quantum Technologies, National University of Singapore, 3 Science drive 2, Singapore 117543\\
$^6$ Department of Physics, National University of Singapore, 2 Science Drive 3, Singapore 117542
}


Correlations cry out for explanation~[1]. Our pre-quantum understanding of correlations relies on a combination of two basic mechanisms. Either the correlated events share a common cause --- such as seeing a flash and hearing the thunder when a lightning strikes --- or one event influences the other --- such as the position of the moon causing the tides. In both cases, we expect the chain of events to satisfy a principle of continuity: that is, the idea that the physical carriers of causal influences propagate continuously through space. In addition, we expect them  --- given the theory of relativity --- to propagate no faster than the speed of light. \\

The experimental violation of Bell inequalities using spacelike separated measurements, however, precludes the explanation of quantum correlations through causal influences propagating at subluminal speed~[2]. Yet, it is always possible, in principle, to explain such experimental violations through models based on hidden influences propagating at a finite speed $v>c$, provided $v$ is large enough~[3,4]. Here~[5], we show that for \emph{any} finite  speed $v>c$, such models predict correlations that can be exploited for faster-than-light communication.
This superluminal communication does not require access to any hidden physical quantities, but only the manipulation of measurement devices at the level of our present-day description of quantum experiments.  Hence, assuming the impossibility of using quantum non-locality for superluminal communication, we exclude any possible explanation of quantum correlations in term of finite-speed influences.

\vfill  

\noindent{\bf References }
\begin{description}
\setlength\itemsep{-3pt}
\item{[1]} A.~Aspect, Nature {\bf 398}, 189 (1999).
\item{[2]} J.~S.~Bell, {\it Speakable and Unspeakable in Quantum Mechanics: Collected papers on quantum philosophy}  (Cambridge University Press, Cambridge, 2004).
\item{[3]} V.~Scarani and N.~Gisin, Phys. Lett. A  {\bf 295}, 167 (2002)
\item{[4]} V.~Scarani and N.~Gisin, Braz. J. Phys.  {\bf 35}, 2A (2005).
\item{[5]} J.-D. Bancal, S. Pironio, A. Ac\'in,  Y.-C. Liang, V. Scarani, and N. Gisin, arXiv:1110.3795 (2011).
\end{description}


%% file: marco/marco.tex
%







\titl{Further investigation on electron stability and 
non-paulian transition in NaI(Tl) crystals} 


\name{ 
R. Bernabei$^{1}$, P. Belli$^{1}$, F. Cappella$^{2}$, R. Cerulli$^{3}$, A. d'Angelo$^{2}$,
C. J. Dai$^{4}$, A. Di Marco$^{1}$, H. L. He$^{4}$, A. Incicchitti$^{2}$, X. H. Ma$^{4}$,
F. Montecchia$^{1,5}$, X. D. Sheng$^{4}$, R.G. Wang$^{4}$, Z.P. Ye$^{4,6}$
}
\adr{ 
$^1$ Dip. di Fisica, Universit\`a di Roma ``Tor Vergata'' and INFN, sez. Roma ``Tor Vergata'', Rome I-00133, Italy\\ 
$^2$ Dip. di Fisica, Universit\`a di Roma ``La Sapienza'' and INFN, sez. Roma ``La Sapienza'', Rome I-00185, Italy\\
$^3$ Laboratori Nazionali del Gran Sasso, I.N.F.N., Assergi, Italy\\
$^4$ IHEP, Chinese Academy, P.O. Box 918/3, Beijing 100039, China\\
$^5$ Laboratorio Sperimentale Policentrico di Ingegneria Medica, Univ. Roma ``Tor Vergata''\\
$^6$ University of Jing Gangshan, Jiangxi, China\\
}
 

A study of non-Paulian (processes normally forbidden by the Pauli Exclusion Principle (PEP)) has been carried out at the Gran 
Sasso National Laboratory by means of the highly radiopure DAMA/LIBRA set-up (sensitive mass of about 250 kg highly 
radiopure NaI(Tl)). In particular, a new improved upper limit for the spontaneous non-Paulian emission rate of protons with 
energy $E_p \le 10$ MeV in $^{23}$Na and $^{127}$I has been obtained: $1.63 \times 10^{33}$ s$^{-1}$ (90\% C.L.). 
The corresponding limit on the relative strength ($\delta^2$) for the searched PEP-forbidden transition is $\delta^2 \lsim (3-4) 
\times 10^{-55}$ (90\% C.L.). PEP-forbidden electron transitions from K-shell in iodine atoms have also 
been investigated and the obtained limit of the lifetime is 
$\tau > 4.7 \times 10^{30}$ s (90\% C.L.), 
corresponding to $\delta^{2}_{e} < 1.28 \times 10^{47}$ (90\% C.L.).This latter limit can also be related to a possible finite size 
of the electron in composite models of quarks and leptons providing superficial violation of the PEP; 
the obtained upper limit on the electron size is $r_0 < 5.7\times 10^{-18}$ cm.

A new search for the possible $e\rightarrow \gamma \nu_{e}$ decay in NaI(Tl) has also been performed;
the analysed exposure is 0.87 ton $\times$ yr.
The obtained limit on the lifetime of the process is: $4 \times 10^{25}$ yr 
(68\% C.L.). It is some orders of magnitude higher 
than the previous best limit available in NaI(Tl) for this decay channel. This result gives for the probability to have 
charge-non-conserving processes $\epsilon^{2} < 2.6 \times 10^{-98}$ (68\% C.L.).

The search for the possible charge-non-conserving electron capture processes with excitation of $^{127}$I at the energy level 418 keV 
has in addition been pursued. The same exposure given above has been considered.
This study has been realized by using the coincidence technique to
reduce the background. The events in coincidence searched for are due to: 
(i) the relaxation of the atomic shells (E$_K \sim$ 33 keV) following the electron disappearance in this process; (ii) the 
de-excitation $\gamma$-emission (E$_\gamma \sim$418 keV). The new limit on the lifetime is: $\tau > 1.2 \times 10^{24}$ yr 
(90\% C.L.). This is the best limit available for NaI(Tl) detectors and it is very close to the best one obtained for $^{129}$Xe.
Stringent restrictions on the relative strengths of the charge-non-conserving processes are derived: $\epsilon^2_W < 
2.5 \times 10^{-25}$; $\epsilon^2_\gamma < 4.95 \times 10^{-39}$.

In the end perspectives for further investigations on CNC and PEP-forbidden transition in DAMA/LIBRA set-up are considered; 
in particular, an estimation on the reachable sensitivity for the lifetime of possible electron disappearance from L-shell 
has been obtained ($\tau \sim 10^{25}$ yr). The electron disappearance due to tunneling-effect 
in extra-dimensional theories has been discussed and at present such experimental sensitivity is very close to the
expected lifetimes in extra-dimensional models with 3 extra dimensions.

\vfill 



%% file: mavromatos/mavromatos.tex

%





\titl{Probing Quantum-Gravity induced Decoherence in (Astro)Particle Physics$^\star$.}

\name{
Nick~E.~Mavromatos$^{1,2}$}

\adr{$^1$ Physics Department, King's College London,  Strand, London WC2R 2LS, UK \\
$^2$ Department of Physics, Theory Division, CERN, CH-1211 Geneva 23, Switzerland } 
$^\star$ {\small \emph{Work supported in part by the London Centre for
Terauniverse Studies (LCTS), using funding from the European Research
Council via the Advanced Investigator Grant 267352}.}


Quantum Gravity (QG) is still an elusive theory, with no experimental signatures so far. Naively, one may think that this had to be expected, because the characteristic mass scale where QG effects set in is the Planck mass, $M_P \sim 1.2. \times 10^{19}$ GeV and thus any low-energy  (compared to Planck scale) searches of QG may be doomed as futile. However, effects associated with QG may show up at much lower energy scales. This is an important feature of all effective field theories in Particle physics that undergo a certain symmetry breakdown, and gravity may not make an exception to the rule.
Indeed, such low energy manifestations of QG may be associated with the breaking of fundamental space-time symmetries, such as Lorentz~[1], or apparent breakdown of unitarity, in the sense that there are degrees of freedom associated with QG that cannot be accessible to a low-energy observer, performing e.g. scattering experiments. This induces decoherence of quantum matter propagating in the background of QG. 
An important aspect of such a decoherence is the ill-defined nature of the CPT operator~[2], which results in measurable (in principle) modifications of the Einstein-Podolsky-Rosen (EPR) correlators of entangled particle states, such as pairs of neutral mesons produced in the appropriate meson factories~[3]. Some estimates, based on stringy QG models entailing low-energy decoherence~[4], indicate that the effect may be falsifiable at the next generation neutral Kaon factories, such as DA$\Phi$NE-2.
Other important effects of this type of CPT violation include different coupling of the QG environment to particles as compared to antiparticles, which may lead to interesting effects for neutrinos in the early universe, namely the possibility that a neutrino-antineutrino asymmetry is created at such early times. The latter is then communicated through Baryon- and Lepton-number violating processes to the Baryon sector of the Standard Model, generating the observed baryon asymmetry in the Universe, without the need for extra sources of CP violation, such as sterile neutrinos~[5]. 

In the talk, I give an overview of various experimental probes of QG decoherence~[6], ranging from the above-mentioned neutral meson systems and factories to cosmic neutrinos. It must be stressed that the modifications of the EPR correlations in entangled particle states, due to intrinsic QG-decoherence-induced CPT violation , seems to be a ``smoking-gun'' type of experiment for such a QG effect, if realised in Nature. 

\vfill  

\noindent{\bf References }
\begin{description}
\setlength\itemsep{-3pt}
\item{[1]} D.~Colladay, V.~A.~Kostelecky,
  Phys.\ Rev.\  {\bf D58 } (1998)  116002;
  V.~A.~Kostelecky,
  Phys.\ Rev.\  {\bf D69 } (2004)  105009.
\item{[2]}  R.~Wald, Phys.\ Rev.\ D{\bf 21} (1980) 2742.
\item{[3]} J.~Bernabeu, N.~E.~Mavromatos and J.~Papavassiliou,
  Phys.\ Rev.\ Lett.\  {\bf 92}  (2004) 131601
\item{[4]} J.~Bernabeu, N.~E.~Mavromatos and S.~Sarkar,
  Phys.\ Rev.\  D {\bf 74} (2006) 045014.
  \item{[5]} G.~Barenboim, N.~E.~Mavromatos,
  Phys.\ Rev.\  {\bf D70 } (2004)  093015.
 \item{[6]} N.~E.~Mavromatos,
  J.\ Phys.\ Conf.\ Ser.\  {\bf 171 } (2009)  012007;
Lect.\ Notes Phys.\  {\bf 669} (2005)  245 and references therein.
\end{description}


%% file: mayburov/mayburov.tex

%





\titl{Randomness and Information Transfer in Quantum Measurements}

\name{
S.~Mayburov$^{1}$, }

\adr{
$^1$ Lebedev Inst. of Physics, Moscow, Russia \\

}


The information transfer in measuring systems (MS) and its
influence on measurement outcome is analysed.
 Formally, any MS can be regarded as
  the information channel between the measured object $S$ and the information
processing and storing system $O$. Hence
  the established fundamental limits on  the amount of information which can be transferred
  via the quantum
  channels can  distort, in principle, the measurement
  outcome [1]. To check this effect, the capacity and
  Holevo limit for some MS models are calculated which, in
  general, confirm its significant scale of information losses.
Earlier it was argued  that due to such information transfer
constraints, the information about the purity of $S$ state
practically can't be transferred to $O$. If this the case, $O$
wouldn't discriminate pure and mixed $S$ ensembles which is
equivalent to the collapse of $S$ pure state [2]. To verify it, we
studied
  the model measurement of $S$ dichotomic  (spin$\frac{1}{2}$) observable $S_z$ performed by MS, which
includes the detector $D$ and $O$, both of them are regarded as
the quantum objects [3].\\

 The measurement of two $S$ ensembles $E_{1,2}$
is compared; $E_1$ includes the pure states which are the
superposition of $S_z$ eigenstates $|S_{1,2}\rangle$ with
amplitudes $a_{1,2}$, another ensemble $E_2$ is the probabilistic
mixture of such eigenstates with the same $\bar{S}_z$. In our
model $S, D, O$  interactions are tuned so that initial
$|S_{1,2}\rangle$ induce the orthogonal $O$ states
$|O_{1,2}\rangle$ with 'pointer' eigenvalues $O_{1,2}$, for
arbitrary pure $S$ state it results in entangled $S, D, O$ state.
   For described  ensembles $S$ purity is
   characterized by the expectation value $\bar \Lambda$ of $S$ observable ${\Lambda}$,
    conjugated to $S_z$; for  $a_1=a_2$ it's given by $\Lambda=S_x$.
     Yet  for our MS final states there is no $O$
    observable $Q$ which expectation value $\bar Q$ depends on
    $\bar \Lambda$. As the result, $O$ can't discriminate the pure and mixed $S$
ensembles with the same $\bar {S}_z$ [3].  The calculation in the
formalism of restrictive maps [3], indicate  that for individual
events these information losses induce the appearance of
randomness in the measurement of $S$ pure ensemble $E_1$, so that
the observed $O$ outcomes correspond to the random $O_{1,2}$
values, i.e. for $O$ the collapse of $S$ pure state occurs [2].
Born rule for outcome probabilities is derived from the linearity
and the invariant properties of MS Hilbert space.
Concerning with MS decogerence effects,
 it's shown  that by itself, due to the unitarity of
decoherence interactions, it can't result in the appearance of
randomness in the measurement outcomes,
 however, it's shown that  MS interactions with its environment
  stabilize the obtained final  MS states
 additionally.



\vfill  

\noindent{\bf References }
\begin{description}
\setlength\itemsep{-3pt}
\item{[1]} B. Schumacher, Phys. Rev.  A {\bf 51} (1995) 2738.
\item{[2]} S. Mayburov,  Int. J. Quant. Inf. {\bf 5} (2007) 279.
\item{[3]} S. Mayburov, Int. J. Quant. Inf., {\bf 9} (2011) 331.
\end{description}


%
%
%
%
%

%% file: mayburov2/mayburov2.tex

%





\titl{Fuzzy Space-time Topology and Gauge Fields}

\name{
S.~Mayburov$^{1}$, }

\adr{
$^1$ Lebedev Inst. of Physics, Moscow, Russia \\

}


Dodson-Zeeman formalism of Fuzzy Geometry  (FG)  is studied as
the possible model of quantum space-time [1].
  FG  elements - fuzzy points (FP)
 $\{a_i\}$,  beside     $a_j \le a_k$ - the standard ordering relation,
  admit also    the  incomparability relation (IR) between them
  - $a_j \sim a_k$, so $\{a_i\}$ set $A^F$ is Poset. In 1-dimensional case,
 Universe is Poset $A^U=A^F \cup X$,
 where $X$ - standard coordinate axe $R^1$.
 $a_i$ properties are detalized by the introduction of
 fuzzy weight $w_i(x)\ge 0$ with norm $1$.
 It supposed that
  FPs $a_i(t) \in A^F$ describe
the  massive particles $m_i$ evolving in time $t$ relative to $X$.
 If  $w_i(x) \ne \delta (x-x_0)$ for arbitrary $x_0$
then $m_i$ coordinate  is principally uncertain [1].\\

IR stipulates the nonboolean weak FG topology, which for $m_i$
free evolution
 induces the nonlinear, quantum-like $w_i(x,t)$ behavior [2].
%
It described as the  phase-like correlations $\alpha(x_1,x_2)$
between $w(x_{1,2},t)$ components. Eventually, $m$ state
$|m(x,t)\}$ is  equivalent to the normalized complex function
$\Psi_t (x)$ in Hilbert space $\cal H$
 for which $|\Psi_t (x)|^2=w(x,t)$. From
the invariance of $|m\}$ evolution  relative to $x,t$ shifts,
 it follows that $\Psi_t (x)$ obeys  QM Schr$\ddot o$dinger equation [2];
the same is true for 3-dimensions. In relativistic FG the minimal
free evolution corresponds to Dirac equation for $m$ with  spin
$\frac {1}{2}$. The minimal  $m_i, m_j$ interactions  can be
introduced as the deformation of $m_{i,j}$ quantum  phases
$\alpha_{i,j}(x,x')$ and are  gauge invariant. In this apporach
the
 interactions of fermion muliplets  are performed by
the corresponding Yang-Mills fields [3].



\vfill  

\noindent{\bf References }
\begin{description}
\setlength\itemsep{-3pt}
\item{[1]} C. Dodson, J. London Math. Soc. {\bf 11} (1975) 465.
\item{[2]} S. Mayburov,  J. Phys A {\bf 41} (2008) 164071.
\item{[3]} S. Mayburov, Int. J. Theor. Phys., {\bf 49} (2010) 3192.
\end{description}






%
%
%
%
%
%

%% file: piacentini/piacentini.tex

%





\titl{INRIM recent results about QM foundations
investigation}

\name{F.~Piacentini$^{1}$}

\adr{$^1$ Optics Division, INRIM, strada delle Cacce 91, 10135 Torino, Italy.}


Quantum Mechanics is one of the most relevant aspects of modern physics, and in the last decades many quantum-based predictions have been experimentally verified, being this theory at the basis of a wide range of research fields (from solid state physics to cosmology and particle physics).
Anyway, many problems related to Quantum Mechanics foundations remain [1], like non-locality, macro-objectivation, wave function reduction, the concept of quantum mechanical measurement etc.
In this proceeding we discuss some specific experiments realized at INRIM about foundations of quantum theory.\\
The first one is a non-classicality test at the single particle level, in two different experimental configurations. The theoretical proposal [2] describes a particular subclass of hidden variable theories in which, having defined two operators $\langle \widehat A\rangle,\langle \widehat B\rangle\geq0$, is always satisfied the inequality:
\begin{equation}
\langle \widehat{A} \rangle \leq \langle
\widehat{B} \rangle \;\;\;\;\Rightarrow\;\;\;\; \langle \widehat{A}^2 \rangle \leq \langle
\widehat{B}^2 \rangle\label{alicki}
\end{equation}
In Quantum Mechanics, instead, it is possible to find such $\langle \widehat A\rangle,\langle \widehat B\rangle$ for which this inequality is violated. After properly defining such operators, we tried to perform this quantumness test exploiting two different heralded single photon sources: with the first one [3], based on Type-I PDC (Parametric Down Conversion) photons generated with a PPLN crystal pumped by a 532 nm CW laser, we obtained a violation of eq. \ref{alicki} of 6 standard deviations ($\sigma$). The successive experimental test was run on PDC photons obtained by a LiIO$_3$ bulk crystal pumped with a 400 nm pulsed laser: here we saw a violation of more than 46 $\sigma$ [4].\\
The second experiment to be presented regards the testing of two restricted local realistic models [5,6] , properly built to describe experiments with entangled photons; the interesting feature of these models is the fact that they are free from the detection loophole, thus they don't rely on the fair sampling assumption. For both models one can obtain a specific inequality (featuring the quantum efficiency of the system itself) that can be violated by Quantum Mechanics.\\
In our experiment [7] we obtained a $|\psi^+\rangle$ Bell state with collinear Type-II PDC photon pairs separated by a non-polarizing beam splitter, and the measurements performed on it did show a clear violation of the inequalities satisfied by these models (11,7 $\sigma$ for the first model [5], and 3,3 $\sigma$ for the second one [6]).\\
In conclusion, these experiments clearly demonstrate that all these HVTs fail to reproduce all the quantum mechanical predictions, being not a valid candidate to replace Quantum Mechanics.

\vfill  

\noindent{\bf References }
\begin{description}
\setlength\itemsep{-3pt}
\item{[1]} M. Genovese, Phys. Rep. \textbf{413} (2005) 319, and references therein.
\item{[2]} R. Alicki and N. van~Ryn, { J. Phys. A} {\bf 41}, (2008) 062001.
\item{[3]} G. Brida {\it et al}, { Opt. Expr.} {\bf 16}, (2008) 11750.
\item{[4]} G. Brida {\it et al}, { PRA} {\bf 79}, (2010) 044102.
\item{[5]} E. Santos, { PLA} {\bf 327}, (2004) 33.
\item{[6]} E. Santos, { EPJD} {\bf 42}, (2007) 501.
\item{[7]} G. Brida, M. Genovese and F. Piacentini, {EPJD} {\bf 44}, (2007) 547.
\end{description}


%% file: rauch/rauch.tex

%





\titl{Hadron interferometry with neutrons}

\name{Helmut Rauch}

\adr{
Atominstitut, Vienna University of Technology, Vienna, Austria
}

Matter wave interferometry provides new information about basic 
features how particles behave in a quantum environment. 
Thermal neutrons with energies about 25 meV are a proper tool for 
such investigations since widely separated coherent beams can be 
produced and manipulated individually. 
Neutrons experience nuclear-, electromagnetic- and gravitational 
interactions and they appear as particles or as waves depending on 
the experimental situation. 
The perfect silicon-crystal interferometer has been used where the 
perfect arrangement of the lattice planes provide coherent beam 
splitting und superposition for such experiments (see figure). 
The 4$\pi$ spinor symmetry of fermions, the 
spin-superposition law and various gravitational effects have been 
verified [1]. 
Topological (geometric) phases can be measured and 
used for advanced wave function reconstruction work. 
Weak and post-selection measurements provide new information about 
the physical situation in the far-field region of a wave packet and 
show how nonlocal interaction effects can be explained by far 
reaching parts of the partial waves that constitute a wave packet. 
Coherence and decoherence are basic features of the wave picture of 
matter, which influence dynamical and topological quantum phases as 
well. 
Related experiments deal with the interaction of neutrons with noisy 
magnetic fields where phase fluctuations and photon-exchange 
processes come into play. 
More recently quantum contextuality has 
been investigated, which indicate an intrinsic entanglement of 
different degrees of freedom [2]. 
The geometric phase shows, surprisingly, a high robustness against 
any fluctuation or dissipation process[3]. 
This may have consequences in quantum communication and 
quantum information systems. 
In a further step Kochen-Specker phenomena have been investigated 
which show stronger correlations of basic features in quantum and 
particle physics than in classical physics.

\begin{figure}[h]
\centering
  \includegraphics[height=.2\textheight]{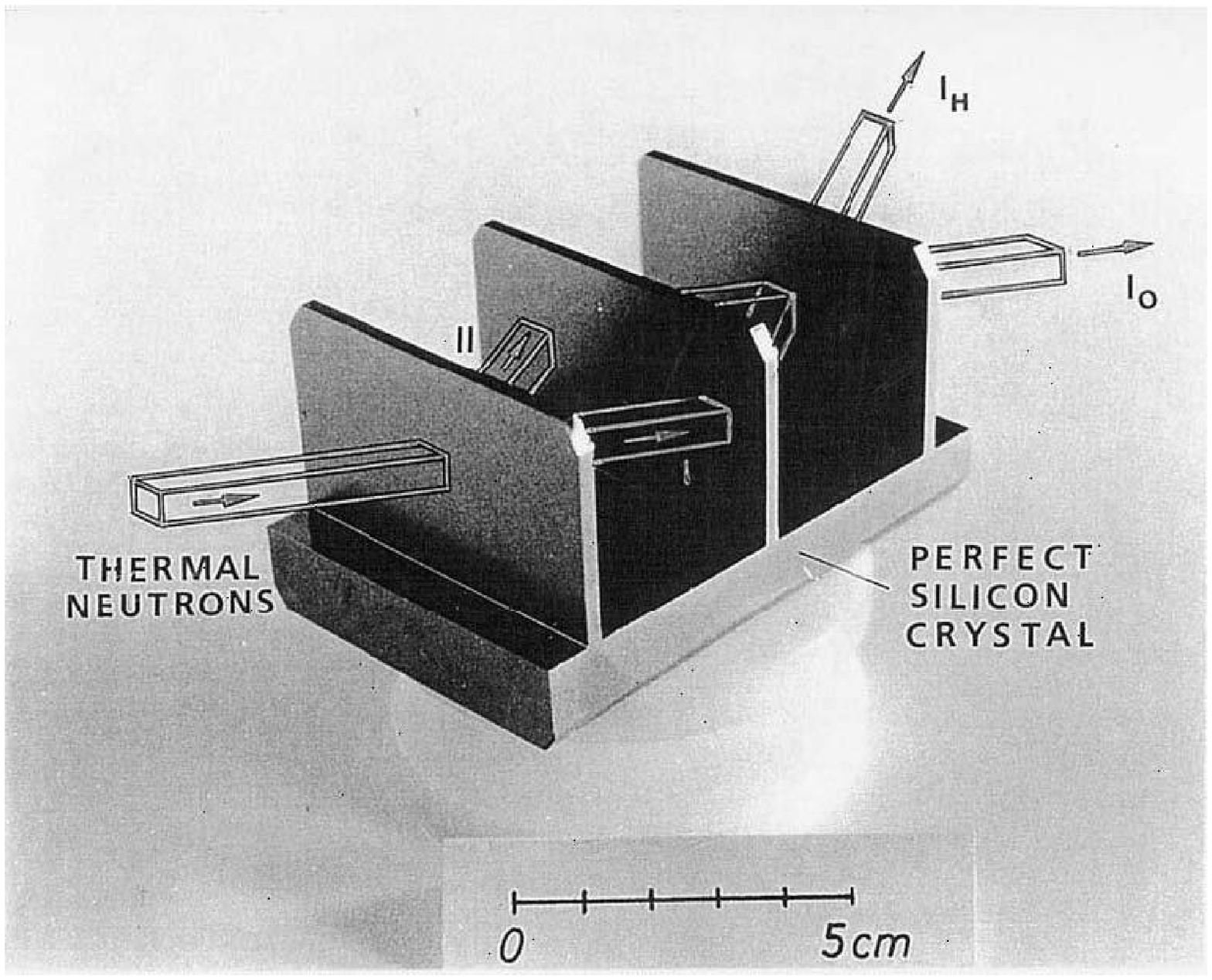}
  \caption{Neutron interferometer}
\end{figure}
\vfill  
\noindent{\bf References }
\begin{description}
\setlength\itemsep{-3pt}
\item{[1]} H.~Rauch and S.~A.~Werner, 
{\it Neutron Interferometry}, Clarendon Press, Oxford 2000. 
\item{[2]} Y.~Hasegawa, R.~Loidl, G.~Badurek, M.~Baron, H.~Rauch, 
Nature {\bf 425} (2003) 45.
\item{[3]} S.~Filipp, J.~Klepp, Y.~Hasegawa, C.~Plonka-Spehr, 
U.~Schmidt, P.~Geltenbort, H.~Rauch, Phys. Rev. Lett. {\bf 102}  
(2009) 030404. 
\end{description}


%% file: rauschenbeutel/rauschenbeutel.tex

%





\titl{Nanofiber Photonics and Quantum Optics}

\name{
A.~Rauschenbeutel$^{1,2}$
}

\adr{
$^1$Vienna Center for Quantum Science and Technology, TU-Wien -- Atominstitut,\\ 
${}~{}$Stadionallee 2, 1020 Vienna, Austria\\
$^2$Institut f\"ur Physik, Johannes Gutenberg-Universit\"at Mainz, 55099 Mainz, Germany
}


We have recently demonstrated a new experimental platform for the simultaneous trapping and optical interfacing of laser-cooled cesium atoms [1]. The scheme uses a multi-color evanescent field surrounding an optical nanofiber to localize the atoms in a one-dimensional optical lattice about 200~nm above the nanofiber surface, see Fig.~1. At the same time, the atoms can be efficiently interrogated with light which is sent through the nanofiber. Remarkably, an ensemble of 2000 trapped atoms almost entirely absorbs a resonant probe field, yielding an optical depth of up to 30, equivalent to an absorbance per atom of 1.5~\%. Moreover, when dispersively interfacing the atoms, we observe a phase shift per atom of $\sim 1$~mrad at a detuning of six times the natural linewidth [2].

Our technique provides unprecedented ease of access for the coherent optical manipulation of laser-cooled neutral atoms and opens the route towards the direct integration of laser-cooled atomic ensembles within fiber networks, an important prerequisite for large scale quantum communication. Moreover, our nanofiber trap is ideally suited to the realization of hybrid quantum systems that combine atoms with solid state quantum devices. Finally, the use of nanofibers for atom trapping allows one to straightforwardly realize intriguing trapping geometries that are not easily accessible with freely propagating laser beams.

\begin{figure}[h]
\centering
  \includegraphics[height=.2\textheight]{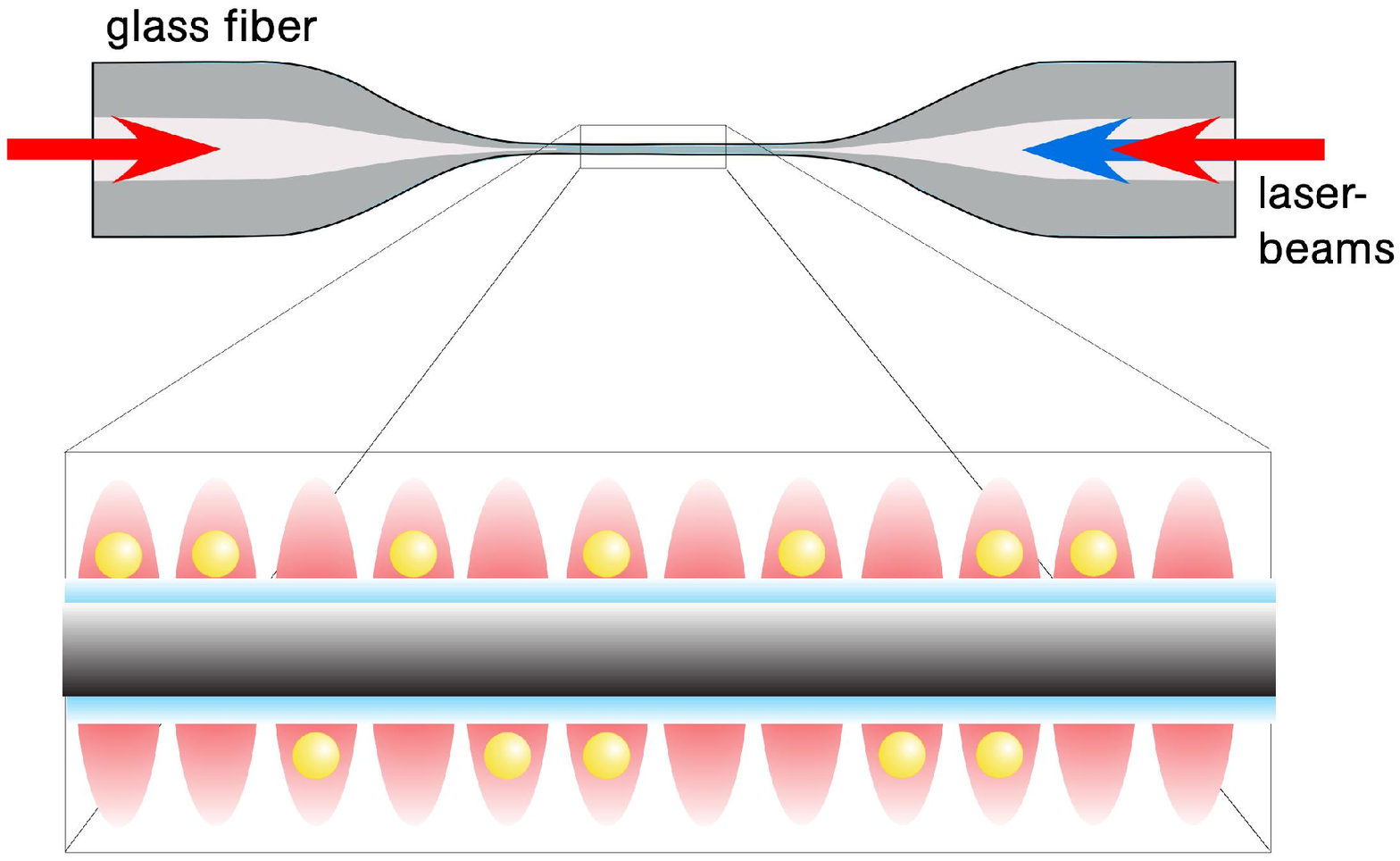}
  \caption{Experimental setup of the fiber-based atom trap. The blue-detuned running wave in combination with the red-detuned standing wave create the trapping potential.}
\end{figure}

\vfill  

\noindent{\bf References }
\begin{description}
\setlength\itemsep{-3pt}
\item{[1]} E. Vetsch, D. Reitz, G. Sagu\'e, R. Schmidt, S. T. Dawkins, and A. Rauschenbeutel, Phys. Rev. Lett. {\bf 104}, 203603 (2010).
\item{[2]} S. T. Dawkins, R. Mitsch, D. Reitz, E. Vetsch, and A. Rauschenbeutel, Phys. Rev. Lett., accepted (2011);
arXiv:1108.2469v2 
\end{description}


%% file: rizzo/rizzo.tex

%





\titl{Spontaneous X-ray emission by free electrons in the collapse models: a closer look}

\name{
A.~Rizzo$^{1}$, C.~Curceanu$^{1}$
}

\adr{
$^1$ Laboratori Nazionali di Frascati, INFN, 00044 Frascati (Rome), Italy \\

}


The Spontaneous Emission by free electrons phenomenon arises from the direct interaction of a free electron and a
fluctuating scalar field, postulated in the framework of Spontaneous Collapse theories as a signature of realisation of the system wave
function collapse which solves the "measurement" problem [1].
The important role played by this new phenomenon in checking such class of theories is easily understandable considering that its cross-section is a function of  $\lambda$, a fundamental 
parameter in the Collapse Master Equations. Nowadays the strongest upper bound on $\lambda$ is set by the $ $experimental search of spontaneous X-ray emitted by free electrons in a low-
energy Ge-based experiment, presented in the work of Q. Fu [2].
We question this result and we present some considerations about the used detectors and the design for a future dedicated experiment.

\begin{equation}
 R(k) = (2.74 \times 10^{-31}) \times 4 \times (8.29 \times 10^{24}) \times(8.6 \times 10^4) \times {1}/{k}\\ 
\end{equation}
 \textit{ \footnotesize{Number of emitted photons for the 4 external Ge electrons in (counts/keV/kg/day). Here k is the emitted photon energy in keV, $ 2.74 \times 10^{-31}=e^2 \lambda / 4 \pi^2 a^2 
 m_{e}^2 $ where  $\lambda = \lambda_{GRW} = 10^{-16} s^{-1}$, $ a= a_{GRW} = 10^{-7} m $ is the correlation length [4].}}\\

Using the formula (1) Fu evaluated the data coming from a first step of the IGEX experiment [3]: he finds a discrepancy between the experimental rate of 0.049 counts/keV/kg/day and the expected 
one of 0.071 counts/kev/kg/day  at the 11 keV (the \textit{anomaly}); he claims for a possible hint to rescale $\lambda_{GRW}$ by a factor 0.45. The hypothesis of a difficulty to scale the data from a 
binning of 100 keV at 11 keV value is questioned, together with other hypoteses [4]. Results of the same analysis performed on data coming 
from  other low-energy Ge-based experiments [4] led us to think about some possible systematics not evaluated in ref. [2]. This hypothesis is actually confirmed in ref [5], where the IGEX 
collaboration claims for a gain stability problem of the detector which affects the data taken during the period considered by Fu.\\
Results of our preliminary analysis on the IGEX corrected data gives an experimental rate 
of 17 and an expected one of 5 counts/keV/kg/day at 11 keV, showing clearly the questionability of the Fu's results. A new and more refined analysis to set the $\lambda$ upper bound is needed.\\
As for a future dedicated experiment, the most promising detector seems to be the PPC, P-type Point Contact Ge: its high sensitivity at lower energies will allow to perform the measurement with 
a small mass ($\sim 500g$) in about one year of exposure, if one considers a data taking in similar conditions as for the CoGent experiment [6].

\vfill  

\noindent{\bf References }
\begin{description}
\setlength\itemsep{-4pt}
\item{[1]} P. Pearle {\it et al.}, Phis. Rev. A . {\bf 39} (1989) 2277.
\item{[2]} Q. Fu, Phys. Rev. A, {\bf 56} (1997) 1806.
\item{[3]} H.S. Miley et al. Phys. Rev. Lett. {\bf 65}, (1990), 3092.
\item{[4]} A. Rizzo, in {\small{\url{https://smimac0.smi.oeaw.ac.at/conference/ect_star_2011/index.html}}}
\item{[5]} C.E. Aalseth et al., IGEX Collaboration, Phys. Rev. C {\bf 59} (1999) 2108.
\item{[6]} P. S. Barbeau et al., J. Cosmol. Astropart. Phys. {\bf 09} (2007) 009.

\end{description}


%% file: vacchini/vacchini.tex

%





\titl{Non-Markovian dynamics: characterizations and measures}

\name{
B.~Vacchini$^{1,2}$
}

\adr{
$^1$ Dipartimento di Fisica, Universit{\`a} degli Studi di
Milano, Via Celoria 16, I-20133 Milan, Italy \\
$^2$ INFN, Sezione di Milano, Via Celoria 16, I-20133
Milan, Italy
}


The description and understanding of non-Markovian dynamics in the
field of open quantum systems is a long standing issue, which has
newly received much attention thanks to proposals for a simple
characterization of non-Markovianity in quantum systems.

Indeed in recent work two measures of non-Markovianity in the quantum framework have been
introduced [1,2], building on two distinct definitions of
Markovian dynamics. In Ref.[1] Markovianity is connected to a
monotonic loss of distinguishability as a function of time for an arbitrary
couple of initial states. This loss of distinguishability is
quantified according to the trace distance. In the approach considered
in Ref.[2] Markovianity is identified with a suitable divisibility property of the mapping
providing the time evolution.

The question about relationship of these two definitions of Markovianity among
themselves and with respect to the classical notion of Markovianity
for a stochastic process then naturally arises. 

In the presentation we have considered in some detail this issue, as
discussed in Ref.[3]. It appears that these two definitions of
non-Markovianity only refer to knowledge of the state, that is to say to
the one-point probability density, so that they cannot directly compare
with the classical definition. Indeed the latter expresses a requirement
on all conditional probability densities of the process, involving
different times and therefore correlations. Nevertheless classical
counterparts of the notion of divisibility of the completely positive
time evolution mapping and of monotonic reduction of the trace
distance among different initial states can be naturally considered,
and can provide sufficient conditions for the detection of
non-Markovianity of a classical process.

To clarify this behavior a class of non-Markovian processes, known as
semi-Markov processes, have been considered and analyzed. Generalizing
the basic idea which characterizes these non-Markovian processes to
the quantum case one can construct a class of completely positive
quantum dynamical evolutions [4], which allow for an explicit analysis
of the two distinct measure of non-Markovianity. It is therefore
possible both to actually evaluate these measures for a suitable
class of dynamics and to point to similarities and differences of the
measures both in the detection and the characterization of
non-Markovianity.

\vfill  

\noindent{\bf References }
\begin{description}
\setlength\itemsep{-3pt}
\item{[1]} H.-P. Breuer, E.-M. Laine, and J.~Piilo, Phys. Rev. Lett. \textbf{103}, (2009) 210401.
\item{[2]} A.~Rivas, S.~F. Huelga, and M.~B. Plenio,
  Phys. Rev. Lett. \textbf{105}, (2010) 050403.
\item{[3]} B.~Vacchini, A.~Smirne, E.-M. Laine, J.~Piilo, and
  H.-P. Breuer, New J. Phys. \textbf{13}, (2011) 093004.
\item{[4]} H.-P. Breuer and B.~Vacchini,
  Phys. Rev. Lett. \textbf{101}, (2008) 140402; Phys. Rev. E \textbf{79}, (2009) 041147.
\end{description}


%% file: vona/vona.tex

%





\titl{Two slits are too many for just one particle}

\name{
N.~Vona$^{1}$
}

\adr{
$^1$ Mathematical Institute of Ludwig-Maximilian University, Munich
}

The double slit interference is probably the most astonishing example among the quantum phenomena.
The effect itself is very simple and general, and applies to any kind of wave.
The astonishing thing is that it applies to single electrons as well [1].
This suggests that the electron is also a wave going through both the slits at the same time, even if it later manifests itself as a particle on the detector.
This double nature of the electron is the keystone of Quantum Mechanics, according to which the electron \emph{is} a wave, behaving like a particle in some special situation.
But if the electron \emph{is} a wave while passing the slits, then it \emph{really} goes through 
two holes at the same time!
Such a phenomenon is so far from our daily experience that if presented with it not knowing anything about Quantum Mechanics we would not believe it to be true.
Is Nature for an electron so different from what it is for us?

Probably most of the people looking back to the experiment without presuming the wave--particle duality would imagine the electron as a particle going through \emph{just one} slit, its motion being later influenced by the other slit, giving rise to the interference pattern on the screen.
This point of view is made rigorous by Bohmian Mechanics, according to which the electron is a point particle driven by a Schr\"odinger wave, that encodes the effect of the second slit.
This theoretical description is much closer to our imagination than the quantum one and therefore is much easier to understand.

The trajectories of the electron predicted by Bohmian Mechanics in the case of a double slit are quite peculiar and it is natural to ask if it is possible to compare this prediction with some experimental result.
To measure a trajectory one needs to measure simultaneously both the position and the velocity of the particle, impeded by Heisemberg principle.
A workaround is to measure them \emph{weakly}, i.e.\ performing non-disturbing measurements that do not change the system appreciably.
The price to pay is a huge uncertainty that makes the single result completely useless, nevertheless the sought information can still be recovered from their statistics.
This approach was recently used in an  experiment with photons, that showed as  a result the characteristic Bohmian pattern [2,3].

Bohmian Mechanics is empirically equivalent to Quantum Mechanics, but much sharper in the formulation and as close to our intuition as a quantum theory could be.
These features allow for a much better understanding of the quantum world than Quantum Mechanics does -- as for Schr\"odinger's cat, the measurement problem, the origin of operators, etc.\ [4,5] -- that could be crucial in the future theoretical developments.

Most importantly,  the electron goes through just one slit!

\vfill  

\noindent{\bf References }
\begin{description}
\setlength\itemsep{-3pt}
\item{[1]} A.\ Tonomura {\it et al.}, Am.\ J.\ Phys.\ {\bf 57} (1989) 117.
\item{[2]} S.\  Kocsis {\it et al.}, Science {\bf 332} (2011) 1170.
\item{[3]} H.M.\ Wiseman, New Journal of Physics {\bf 9} (2007) 165.
\item{[4]} D.\ D\"urr, S.\ Goldstein, N.\ Zangh\`i, Quantum Physics without Quantum Philosophy, Springer (2012).
\item{[5]} D.\ D\"urr, S.\ Teufel, Bohmian mechanics: the physics and mathematics
of quantum theory, Springer (2009).
\end{description}


%% file: widmann/widmann.tex

%





\titl{Testing $\mathcal{CPT}$ with antiprotonic helium and antihydrogen}

\name{
E.~Widmann
}

\adr{
Stefan Meyer Institute, 1090 Vienna, Austria}


$\mathcal{CPT}$ symmetry states that a system is invariant under the combined transformations $\mathcal{C}$ (charge conjugation, \emph{i.e.} exchanging a particle by its antiparticle), parity $\mathcal{P}$ and time reversal $\mathcal{T}$. This $\mathcal{CPT}$ theorem is the consequence of certain properties of the quantum field theories used in the Standard Model of particle physics (SM). Testing $\mathcal{CPT}$ symmetry is thus a test of the validity of the basis of the SM, which has been very successful in explaining most of the observations in particle physics, but is incomplete. Extensions of the SM like string theory do not have the properties required by the mathematical proof of the $\mathcal{CPT}$ theorem -- like point-like particles -- and may as a consequence lead to $\mathcal{CPT}$ violations. 
Precise measurements of properties of particles and antiparticles are very sensitive tests of $\mathcal{CPT}$, and the spectroscopy of exotic atoms containing antiprotons are among the most promising ways to reach the highest precision. 

{\bf Antiprotonic helium} ($\overline{\mathrm p}$He$^{+}$) is a three-body system consisting of a helium nucleus, an electron, and an antiproton. It comprises a series of highly excited states that have lifetimes in the micro-second range making it accessible to laser and microwave spectroscopy [1]. 
Comparing the observed transition energies to three-body QED calculations, the mass $(m_{\overline{p}}-m_{p})/m_{av}< 7\times10^{-10}$ [2] and magnetic moment ($\mu_{p}-\mu_{\overline{p}})/\mu_{p} < 2.9\times10^{-3}$ [3] of the antiproton could be determined to the highest precision, thus creating some of the most sensitive tests of test $\mathcal{CPT}$ in the hadron sector. 

{\bf Antihydrogen} $\overline{\mathrm H} \equiv \overline{\mathrm{p}}$e$^{+}$, the simplest atom consisting of pure antimatter, offers very high precision tests of $\mathcal{CPT}$ by comparing transitions to well-measured counterparts of hydrogen. The ASACUSA collaboration has proposed to measure the ground-state hyperfine splitting (GS-HFS) in antihydrogen [4], whose equivalent has been measured to $\sim 10^{-12}$ relative precision in the hydrogen maser. The GS-HFS is caused by the spin-spin interaction of the antiproton and positron, and below a level of $\sim 10^{-5}$ is also sensitive to the magnetic radius of the (anti)proton, \emph{i.e.} their hadronic structure. 

The experimental setup which was tested for the first time during November 2011 makes use of an atomic beam line using inhomogeneous magnetic fields, similar to how the hydrogen hyperfine structure was measured by Rabi and others in the 1930s to 1950s. The necessary polarized beam of antihydrogen atoms will be produced by a novel magnetic trap configuration called ``cusp'' trap. In 2010 a major break through was achieved when for the first time the formation of antihydrogen was shown in the cusp trap [5]. In the first stage of experiments, using this setup it will be possible to measure the GS-HFS of antihydrogen to $\sim 10^{-7}$ relative precision. An improvement of one order of magnitude can be achieved in a second phase by using the separated oscillatory field method of Ramsey. The ultimate resolution will be possible using an atomic fountain of laser-cooled antihydrogen atoms, an established technique for ordinary atoms.

\vfill  

\noindent{\bf References }
\begin{description}
\setlength\itemsep{-3pt}
\item{[1]} R.~Hayano, M.~Hori, D.~Horv{\'a}th, and E.~Widmann, Rep. Prog.
  Phys. \textbf{70} (2007) 1995.
\item{[2]} M.~Hori {\it et al.}, Nature \textbf{475} (2011) 484.
\item{[3]} T.~Pask {\it et al.}, Physics Letters B \textbf{678}(2009) 55.
\item{[4]} E.~Widmann {\it et al.}, CERN-SPSC-2003-009, CERN, Geneva, Switzerland (2003).
\item{[4]}  Y.~Enomoto {\it et al.}, Phys. Rev. Lett. \textbf{105} (2010) 243401.
\end{description}


%% file: zavattini/zavattini.tex

%





\titl{ Measuring the magnetic birefringence of vacuum: status and perspectives of the PVLAS experiment}

\name{
G.~Zavattini$^{1}$, F.~Della~Valle$^{2}$, G.~Di~Domenico$^{1}$, U.~Gastaldi$^{3}$, E.~Milotti$^{2}$, R.~Pengo$^{3}$, L.~Piemontese$^{1}$, and G.~Ruoso$^{3}$
}

\adr{
$^1$ Department of Physics, University of Ferrara and INFN-Ferrara, Ferrara, Italy \\
$^2$ Department of Physics, University of Trieste and INFN-Trieste, Trieste, Italy \\
$^3$ INFN-Laboratori Nazionali di Legnaro, Legnaro, Italy \\
}


It is well known that Maxwell's equations in vacuum are linear: the superposition principle is valid for electromagnetic fields. Heisenberg's principle, though, allows vacuum to fluctuate producing virtual pairs of particles which, if charged, lead to non linear effects in vacuum such as light-light scattering through the box diagram. Furthermore, hypothetical neutral particles coupling to two photons may also allow the interaction between two photons. These effects result in $v \ne c$, birefringence, and dichroism in the presence of an external field (magnetic or electric). The PVLAS experiment, with a new setup being currently mounted at the Physics Department of the University of Ferrara, Ferrara, Italy, has the goal of measuring both the real and imaginary part of the index of refraction of light when traversing a magnetic field. With a previous version [1] of the PVLAS apparatus the best limits on both vacuum magnetic birefringence and dichroism were set. From these measurements we also extracted the best limits on the light - light total cross section: $\sigma_{\gamma\gamma} < 4.6\cdot 10^{-62}$ m$^{2}$ at $\lambda = 1064$ nm.

Without having to invoke new physics, QED predicts such non linear effects which have never been directly observed yet. In particular a magnetic field will generate a birefringence $\Delta n_{\rm QED} = 3 A_{e} B^{2}$ where $A_{e}=\frac{2}{45\mu_{\rm 0}}\frac{\alpha^{2}\mathchar'26\mkern-10mu\lambda_e^{3}}{m_{e}c^{2}}=1.32\cdot10^{-24}$ T$^{-2}$. With an external field strength of $B = 2.5$ T, $\Delta n_{\rm QED} = 2.5\cdot 10^{-23}$.

The new apparatus is a sensitive ellipsometer based on a Fabry-Perot cavity with very high finesse ($F > 2\cdot 10^{5}$), two identical rotating 2.5 T dipole permanent magnets each of length $L = 1$ m, and an ellipticity modulator for heterodyne detection. A scheme of the measurement principle is shown in Figure 1. In our configuration the ellipticity induced by the vacuum magnetic birefringence is $\psi = \frac{2F L \Delta n_{\rm QED}}{\lambda} = 1.9\cdot 10^{-11}$. 

\begin{figure}[h]
\centering
  \includegraphics[height=.1\textheight]{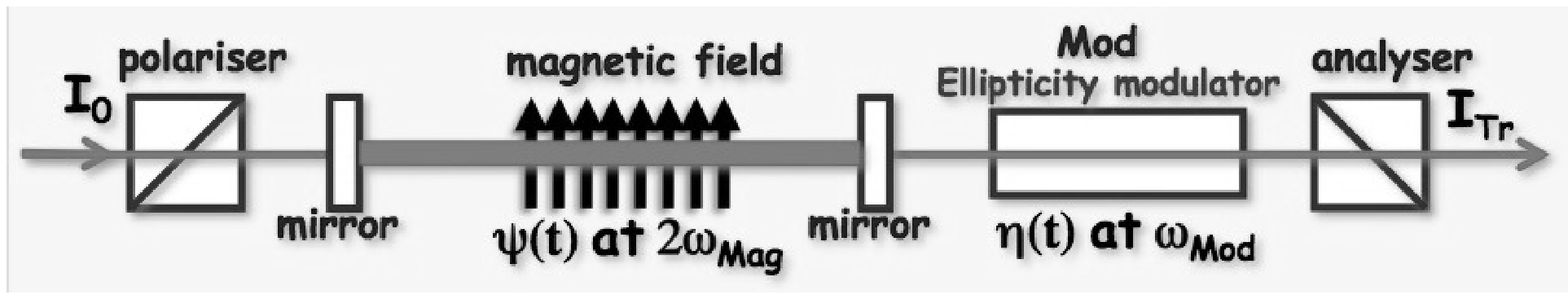}
  \caption{Principle scheme of the measurement. The Fabry-Perot cavity increases the induced ellipticity $\psi$ by a factor proportional to the finesse $F$ whereas the rotating magnets  make this ellipticity time dependent. The ellipticity $\psi(t)$ then beats with the modulator's ellipticity. The  ellipticity amplitude is then extracted from a Fourier analysis of $I_{\rm Tr}$.}
\end{figure}

At present we are running a smaller test apparatus for studying systematic errors. The ellipticity sensitivity is $s = 3 \cdot 10^{-7} 1/\sqrt{\rm Hz}$ with a finesse $F = 245000$.

\vfill  

\noindent{\bf References }
\begin{description}
\setlength\itemsep{-3pt}
\item{[1]} M. Bregant {\it et al.}, Physical Review D {\bf 78} (2008) 032006 and E. Zavattini {\it et al.}, Physical Review D {\bf 77} (2008) 032006.
\end{description}


%% file: program/program.tex

%






\vspace{5cm}
\begin{center}
{\Huge ECT*}\\
\vspace{1cm}
{\Large EUROPEAN CENTRE FOR THEORETICAL STUDIES}\\
{\Large IN NUCLEAR PHYSICS AND RELATED AREAS}\\
{\large \textbf{Trento}}{\large , ITALY}\\

\vspace{5cm}
{\large Speakable in quantum mechanics: }\\
{\large atomic, nuclear and subnuclear physics tests}

\vspace{2cm}
{\large ECT*-Trento, 29 August - 2 September, 2011}

\vspace{110pt}
{\large \textit{Organized by}}\\
{\large \textit{Catalina Curceanu  - LNF-INFN, Frascati (Italy)}}\\
{\large \textit{Johann Marton -- SMI-Vienna (Austria)}}\\
{\large \textit{Edoardo Milotti -- Universita' di Trieste and INFN Trieste}}
\end{center}

\newpage
\begin{longtable}{|p{4.500in}|}
\hline
\begin{minipage}[t]{4.500in}\centering Monday 29 August\end{minipage}\\
\hline
\end{longtable}

\vspace{13pt}
\textit{Morning Session. }\textbf{Chair J. Marton}

\begin{longtable}{lll}
\hline
\multicolumn{1}{|p{1.288in}|}{\begin{minipage}[t]{1.288in}\raggedright
9:00 -- 9:45\end{minipage}} & \multicolumn{2}{p{3.212in}|}{\begin{minipage}[t]{3.212in}\raggedright
REGISTRATION\end{minipage}}\\
\hline
\multicolumn{1}{|p{1.288in}|}{\begin{minipage}[t]{1.288in}\raggedright
9:45 -- 10:00\end{minipage}} & \multicolumn{1}{p{0.912in}|}{\begin{minipage}[t]{0.912in}\raggedright
C. Curceanu\end{minipage}} & \multicolumn{1}{p{2.300in}|}{\begin{minipage}[t]{2.300in}\raggedright
\textit{Welcome address}\end{minipage}}\\
\hline
\multicolumn{1}{|p{1.288in}|}{\begin{minipage}[t]{1.288in}\raggedright
10:00 -- 10:40\end{minipage}} & \multicolumn{1}{p{0.912in}|}{\begin{minipage}[t]{0.912in}\raggedright
N. Mavromatos\end{minipage}} & \multicolumn{1}{p{2.300in}|}{\begin{minipage}[t]{2.300in}\raggedright
\textit{Probing Quantum-Gravity-induced Decoherence in (Astro) Particle Physics}\end{minipage}}\\
\hline
\multicolumn{1}{|p{1.288in}|}{\begin{minipage}[t]{1.288in}\raggedright
\textit{10:40 -- 11:10}\end{minipage}} & \multicolumn{1}{p{0.912in}|}{\begin{minipage}[t]{0.912in}\raggedright
\textit{Coffee break}\end{minipage}} & \multicolumn{1}{p{2.300in}|}{\begin{minipage}[t]{2.300in}\raggedright
\end{minipage}}\\
\hline
\multicolumn{1}{|p{1.288in}|}{\begin{minipage}[t]{1.288in}\raggedright
11:10 -- 11:50\end{minipage}} & \multicolumn{1}{p{0.912in}|}{\begin{minipage}[t]{0.912in}\raggedright
A. di Domenico \end{minipage}} & \multicolumn{1}{p{2.300in}|}{\begin{minipage}[t]{2.300in}\raggedright
\textit{CPT symmetry, quantum mechanics and neutral kaons}\end{minipage}}\\
\hline
\multicolumn{1}{|p{1.288in}|}{\begin{minipage}[t]{1.288in}\raggedright
11:50 -- 12:30\end{minipage}} & \multicolumn{1}{p{0.912in}|}{\begin{minipage}[t]{0.912in}\raggedright
B. Hiesmayr\end{minipage}} & \multicolumn{1}{p{2.300in}|}{\begin{minipage}[t]{2.300in}\raggedright
\textit{Bell's [Un]speakable in the Neutral Kaon System}\end{minipage}}\\
\hline
\multicolumn{1}{|p{1.288in}|}{\begin{minipage}[t]{1.288in}\raggedright
\textit{12:30 -- 14.20}\end{minipage}} & \multicolumn{1}{p{0.912in}|}{\begin{minipage}[t]{0.912in}\raggedright
\textit{Lunch}\end{minipage}} & \multicolumn{1}{p{2.300in}|}{\begin{minipage}[t]{2.300in}\raggedright
\end{minipage}}\\
\hline
\end{longtable}

\vspace{27pt}
\textit{Afternoon Session. }\textbf{Chair F. de Martini}

\begin{longtable}{|p{1.288in}|p{1.079in}|p{2.133in}|}
\hline
\begin{minipage}[t]{1.288in}\raggedright 14:20 -- 15:00\end{minipage} & \begin{minipage}[t]{1.079in}\raggedright R. 
Ursin/ N. Kiesel\end{minipage} & \begin{minipage}[t]{2.133in}\raggedright \textit{Tests 
of quantum mechanics using entangled photons}\end{minipage}\\
\hline
\begin{minipage}[t]{1.288in}\raggedright 15:00 -- 15:40\end{minipage} & \begin{minipage}[t]{1.079in}\raggedright A. 
Rauschenbeutel\end{minipage} & \begin{minipage}[t]{2.133in}\raggedright \textit{Nanofiber 
Photonics and Quantum Optics}\end{minipage}\\
\hline
\begin{minipage}[t]{1.288in}\raggedright \textit{15:40 -- 16:10}\end{minipage} & \begin{minipage}[t]{1.079in}\raggedright \textit{Coffee 
break}\end{minipage} & \begin{minipage}[t]{2.133in}\raggedright \end{minipage}\\
\hline
\begin{minipage}[t]{1.288in}\raggedright 16:10 -- 16:50\end{minipage} & \begin{minipage}[t]{1.079in}\raggedright F. 
Piacentini\end{minipage} & \begin{minipage}[t]{2.133in}\raggedright \textit{INRIM 
recent results about QM foundations investigation}\end{minipage}\\
\hline
\end{longtable}

\vspace{13pt}
\textit{EU Project:}

\begin{longtable}{|p{1.288in}|p{1.079in}|p{2.133in}|}
\hline
\begin{minipage}[t]{1.288in}\raggedright 16:50 -- 17:30\end{minipage} & \begin{minipage}[t]{1.079in}\raggedright A. 
Bassi\end{minipage} & \begin{minipage}[t]{2.133in}\raggedright COST EU Project 
-- presentation and discussions\end{minipage}\\
\hline
\end{longtable}

\newpage
\begin{longtable}{|p{4.500in}|}
\hline
\begin{minipage}[t]{4.500in}\centering Tuesday 30 August\end{minipage}\\
\hline
\end{longtable}

\vspace{13pt}
\textit{Morning Session. }\textbf{Chair A. di Domenico}

\vspace{12pt}
\begin{longtable}{|p{1.288in}|p{0.912in}|p{2.300in}|}
\hline
\begin{minipage}[t]{1.288in}\raggedright 9:00 -- 9:40\end{minipage} & \begin{minipage}[t]{0.912in}\raggedright A. 
Bassi \end{minipage} & \begin{minipage}[t]{2.300in}\raggedright \textit{Is Quantum 
Theory exact? Collapse models and the possibility of a break down of quantum mechanics 
towards the macroscopic scale}\end{minipage}\\
\hline
\begin{minipage}[t]{1.288in}\raggedright 9:40 -- 10:20\end{minipage} & \begin{minipage}[t]{0.912in}\raggedright L. 
Ferialdi\end{minipage} & \begin{minipage}[t]{2.300in}\raggedright \textit{Non-markovian 
features in quantum dynamics}\end{minipage}\\
\hline
\begin{minipage}[t]{1.288in}\raggedright \textit{10:20 -- 10:50}\end{minipage} & \begin{minipage}[t]{0.912in}\raggedright \textit{coffee 
break}\end{minipage} & \begin{minipage}[t]{2.300in}\raggedright \end{minipage}\\
\hline
\begin{minipage}[t]{1.288in}\raggedright 10:50 -- 11:30\end{minipage} & \begin{minipage}[t]{0.912in}\raggedright Y.-C. 
Liang\end{minipage} & \begin{minipage}[t]{2.300in}\raggedright \textit{Finite-speed 
hidden influences are incompatible with quantum theory}\end{minipage}\\
\hline
\begin{minipage}[t]{1.288in}\raggedright 11:30 -- 12:10\end{minipage} & \begin{minipage}[t]{0.912in}\raggedright G.M. 
d'Ariano\end{minipage} & \begin{minipage}[t]{2.300in}\raggedright \textit{A Quantum-Digital 
Universe: a Quantum Cellular Automata Approajh to Field Theory}\end{minipage}\\
\hline
\begin{minipage}[t]{1.288in}\raggedright 12:10 -- 12:50\end{minipage} & \begin{minipage}[t]{0.912in}\raggedright F. 
de Martini \end{minipage} & \begin{minipage}[t]{2.300in}\raggedright \textit{An 
initio derivation of Dirac's equation by conformal geometry: the Affine Quantum 
Mechanics}\end{minipage}\\
\hline
\begin{minipage}[t]{1.288in}\raggedright 12:50 -- 14:20\end{minipage} & \begin{minipage}[t]{0.912in}\raggedright Lunch\end{minipage} & \begin{minipage}[t]{2.300in}\raggedright \end{minipage}\\
\hline
\end{longtable}

\vspace{24pt}
\textit{Afternoon Session. }\textbf{Chair B. Hiesmayr}

\vspace{12pt}

\begin{longtable}{|p{1.288in}|p{0.912in}|p{2.300in}|}
\hline
\begin{minipage}[t]{1.288in}\raggedright 14:20 -- 15:00\end{minipage} & \begin{minipage}[t]{0.912in}\raggedright S. 
Gerlich\end{minipage} & \begin{minipage}[t]{2.300in}\raggedright \textit{Recent 
developments in molecule interferometry}\end{minipage}\\
\hline
\begin{minipage}[t]{1.288in}\raggedright 15:00 -- 15:40\end{minipage} & \begin{minipage}[t]{0.912in}\raggedright N. 
Kiesel\end{minipage} & \begin{minipage}[t]{2.300in}\raggedright \textit{Quantum-optomechanics: 
Quantum Control of Massive Objects}\end{minipage}\\
\hline
\begin{minipage}[t]{1.288in}\raggedright \textit{15:40 -- 16.10}\end{minipage} & \begin{minipage}[t]{0.912in}\raggedright \textit{Coffee 
break}\end{minipage} & \begin{minipage}[t]{2.300in}\raggedright \end{minipage}\\
\hline
\begin{minipage}[t]{1.288in}\raggedright 16:10 -- 16:50\end{minipage} & \begin{minipage}[t]{0.912in}\raggedright H. 
Rauch\end{minipage} & \begin{minipage}[t]{2.300in}\raggedright \textit{Hadron Interferometry 
with Neutrons}\end{minipage}\\
\hline
\begin{minipage}[t]{1.288in}\raggedright 16:50 -- 17:30\end{minipage} & \begin{minipage}[t]{0.912in}\raggedright G. 
Zavattini\end{minipage} & \begin{minipage}[t]{2.300in}\raggedright \textit{Measuring 
the magnetic birefringence of vacuum: status and future perspectives of the PVLAS 
experiment}\end{minipage}\\
\hline
\end{longtable}

\newpage
\begin{longtable}{|p{4.500in}|}
\hline
\begin{minipage}[t]{4.500in}\centering Wednesday 31 August\end{minipage}\\
\hline
\end{longtable}

\vspace{73pt}
\textit{Morning Session. }\textbf{Chair H. Rauch}

\vspace{12pt}
\begin{longtable}{|p{1.288in}|p{0.912in}|p{2.300in}|}
\hline
\begin{minipage}[t]{1.288in}\raggedright 9:00 -- 9:40\end{minipage} & \begin{minipage}[t]{0.912in}\raggedright P. 
Belli\end{minipage} & \begin{minipage}[t]{2.300in}\raggedright \textit{The Dark 
Matter annual modulation signature as a probe of the dark side of the Universe}\end{minipage}\\
\hline
\begin{minipage}[t]{1.288in}\raggedright 9:40 -- 10:20\end{minipage} & \begin{minipage}[t]{0.912in}\raggedright A. 
di Marco\end{minipage} & \begin{minipage}[t]{2.300in}\raggedright \textit{Further 
investigation of electron stability and non-paulian transition in NaI(Tl) crystals}\end{minipage}\\
\hline
\begin{minipage}[t]{1.288in}\raggedright \textit{10:20 -- 10:50}\end{minipage} & \begin{minipage}[t]{0.912in}\raggedright \textit{coffee 
break}\end{minipage} & \begin{minipage}[t]{2.300in}\raggedright \end{minipage}\\
\hline
\begin{minipage}[t]{1.288in}\raggedright 10:50 -- 11:30\end{minipage} & \begin{minipage}[t]{0.912in}\raggedright K. 
Formenko\end{minipage} & \begin{minipage}[t]{2.300in}\raggedright \textit{Study 
of the rare processes with the BOREXINO detector}\end{minipage}\\
\hline
\begin{minipage}[t]{1.288in}\raggedright 11:30 -- 12:10\end{minipage} & \begin{minipage}[t]{0.912in}\raggedright C. 
Curceanu \end{minipage} & \begin{minipage}[t]{2.300in}\raggedright \textit{From 
the Pauli Exclusion principle violation tests (VIP experiment) to collapse models 
experimental investigation plans}\end{minipage}\\
\hline
\begin{minipage}[t]{1.288in}\raggedright 12:10 -- 12:50\end{minipage} & \begin{minipage}[t]{0.912in}\raggedright S. 
Mayburov\end{minipage} & \begin{minipage}[t]{2.300in}\raggedright \textit{Randomness 
and Information Transfer in Quantum Mechanics}\end{minipage}\\
\hline
\begin{minipage}[t]{1.288in}\raggedright 12:50 -- 14:20\end{minipage} & \begin{minipage}[t]{0.912in}\raggedright Lunch\end{minipage} & \begin{minipage}[t]{2.300in}\raggedright \end{minipage}\\
\hline
\end{longtable}

\vspace{36pt}
\textit{Afternoon Reserved for Discussions}

\vspace{55pt}
SOCIAL DINNER

\newpage
\begin{longtable}{|p{4.500in}|}
\hline
\begin{minipage}[t]{4.500in}\centering Thursday 1 September\end{minipage}\\
\hline
\end{longtable}

\vspace{13pt}
\textit{Morning Session. }\textbf{ Chair N. Zanghi}

\begin{longtable}{|p{1.288in}|p{0.912in}|p{2.300in}|}
\hline
\begin{minipage}[t]{1.288in}\raggedright 9:30 -- 10:10\end{minipage} & \begin{minipage}[t]{0.912in}\raggedright B. 
Dakic\end{minipage} & \begin{minipage}[t]{2.300in}\raggedright \textit{Probabilistic 
theories: classical, quantum and beyond quantum}\end{minipage}\\
\hline
\begin{minipage}[t]{1.288in}\raggedright 10:10 -- 10:50\end{minipage} & \begin{minipage}[t]{0.912in}\raggedright R. 
Floreanini\end{minipage} & \begin{minipage}[t]{2.300in}\raggedright \textit{Quantum 
contextuality in systems of identical particles}\end{minipage}\\
\hline
\begin{minipage}[t]{1.288in}\raggedright \textit{10:50 -- 11.20}\end{minipage} & \begin{minipage}[t]{0.912in}\raggedright \textit{Coffee 
break}\end{minipage} & \begin{minipage}[t]{2.300in}\raggedright \end{minipage}\\
\hline
\begin{minipage}[t]{1.288in}\raggedright 11:20 -- 12:00\end{minipage} & \begin{minipage}[t]{0.912in}\raggedright A. 
Hayrapetyan\end{minipage} & \begin{minipage}[t]{2.300in}\raggedright \textit{Probability 
amplitudes of two-levels atoms beyond the dipole approximation}\end{minipage}\\
\hline
\begin{minipage}[t]{1.288in}\raggedright 12:00 -- 12:40\end{minipage} & \begin{minipage}[t]{0.912in}\raggedright K. 
Jungmann/\linebreak
L. Willmann\end{minipage} & \begin{minipage}[t]{2.300in}\raggedright \textit{C, 
P and T and Lorentz Invariance with Radioactive Atoms}\end{minipage}\\
\hline
\begin{minipage}[t]{1.288in}\raggedright \textit{12:40 -- 14:20}\end{minipage} & \begin{minipage}[t]{0.912in}\raggedright \textit{Lunch}\end{minipage} & \begin{minipage}[t]{2.300in}\raggedright \end{minipage}\\
\hline
\end{longtable}

\vspace{12pt}
\textit{Afternoon Session. }\textbf{Chair E. Milotti}

\vspace{13pt}
\begin{longtable}{|p{1.288in}|p{0.912in}|p{2.300in}|}
\hline
\begin{minipage}[t]{1.288in}\raggedright 14:20 -- 15:00\end{minipage} & \begin{minipage}[t]{0.912in}\raggedright B. 
Vacchini\end{minipage} & \begin{minipage}[t]{2.300in}\raggedright \textit{Non-markovian 
dynamics: characterization and measures}\end{minipage}\\
\hline
\begin{minipage}[t]{1.288in}\raggedright 15:00 -- 15:40\end{minipage} & \begin{minipage}[t]{0.912in}\raggedright N. 
Vona\end{minipage} & \begin{minipage}[t]{2.300in}\raggedright \textit{Two slits 
are too many for just one particle}\end{minipage}\\
\hline
\begin{minipage}[t]{1.288in}\raggedright \textit{15:40 -- 16.10}\end{minipage} & \begin{minipage}[t]{0.912in}\raggedright \textit{Coffee 
break}\end{minipage} & \begin{minipage}[t]{2.300in}\raggedright \end{minipage}\\
\hline
\begin{minipage}[t]{1.288in}\raggedright 16:10 -- 16:50\end{minipage} & \begin{minipage}[t]{0.912in}\raggedright N. 
Zanghi\end{minipage} & \begin{minipage}[t]{2.300in}\raggedright \textit{Speakable 
(without unspeakable) in Bohmian mechanics}\end{minipage}\\
\hline
\begin{minipage}[t]{1.288in}\raggedright 16:50 -- 17:30\end{minipage} & \begin{minipage}[t]{0.912in}\raggedright A. 
Diaz-Torres\end{minipage} & \begin{minipage}[t]{2.300in}\raggedright \textit{Quantum 
decoherence in low-energy nuclear reaction dynamics}\end{minipage}\\
\hline
\end{longtable}

\newpage
\begin{longtable}{|p{4.500in}|}
\hline
\begin{minipage}[t]{4.500in}\centering Friday 2 September\end{minipage}\\
\hline
\end{longtable}

\vspace{13pt}
\textit{Morning Session. }\textbf{ Chair J. Marton}

\vspace{27pt}
\begin{longtable}{lll}
\hline
\multicolumn{1}{|p{1.288in}|}{\begin{minipage}[t]{1.288in}\raggedright
9:00 -- 9:40\end{minipage}} & \multicolumn{1}{p{0.912in}|}{\begin{minipage}[t]{0.912in}\raggedright
S. Mayburov\end{minipage}} & \multicolumn{1}{p{2.300in}|}{\begin{minipage}[t]{2.300in}\raggedright
\textit{Fuzzy space-time topology and the nature of quantization}\end{minipage}}\\
\hline
\multicolumn{1}{|p{1.288in}|}{\begin{minipage}[t]{1.288in}\raggedright
9:40 -- 10:20\end{minipage}} & \multicolumn{1}{p{0.912in}|}{\begin{minipage}[t]{0.912in}\raggedright
M. Gouanere \end{minipage}} & \multicolumn{1}{p{2.300in}|}{\begin{minipage}[t]{2.300in}\raggedright
\textit{A Search for the de Broglie Particle Internal Clock by Means of Electron 
Channeling }\end{minipage}}\\
\hline
\multicolumn{1}{|p{1.288in}|}{\begin{minipage}[t]{1.288in}\raggedright
\textit{10:20 -- 10:50}\end{minipage}} & \multicolumn{1}{p{0.912in}|}{\begin{minipage}[t]{0.912in}\raggedright
\textit{coffee break}\end{minipage}} & \multicolumn{1}{p{2.300in}|}{\begin{minipage}[t]{2.300in}\raggedright
\end{minipage}}\\
\hline
\multicolumn{1}{|p{1.288in}|}{\begin{minipage}[t]{1.288in}\raggedright
10:50 -- 11:30\end{minipage}} & \multicolumn{1}{p{0.912in}|}{\begin{minipage}[t]{0.912in}\raggedright
E. Widmann\end{minipage}} & \multicolumn{1}{p{2.300in}|}{\begin{minipage}[t]{2.300in}\raggedright
\textit{Testing CTP symmetry with antiprotonic helium and antihydrogen}\end{minipage}}\\
\hline
\multicolumn{1}{|p{1.288in}|}{\begin{minipage}[t]{1.288in}\raggedright
11:30 -- 12:00\end{minipage}} & \multicolumn{1}{p{0.912in}|}{\begin{minipage}[t]{0.912in}\raggedright
A. Clozza\end{minipage}} & \multicolumn{1}{p{2.300in}|}{\begin{minipage}[t]{2.300in}\raggedright
\textit{Future plans for upgrade of the VIP experiment for the study of the violation 
of the Pauli Exclusion Principle for electrons}\end{minipage}}\\
\hline
\multicolumn{1}{|p{1.288in}|}{\begin{minipage}[t]{1.288in}\raggedright
12:00 -- 12:30\end{minipage}} & \multicolumn{1}{p{0.912in}|}{\begin{minipage}[t]{0.912in}\raggedright
A. Rizzo\end{minipage}} & \multicolumn{1}{p{2.300in}|}{\begin{minipage}[t]{2.300in}\raggedright
\textit{Spontaneous X-ray emission by free electrons in collapse models: a closer 
look}\end{minipage}}\\
\hline
\multicolumn{1}{|p{1.288in}|}{\begin{minipage}[t]{1.288in}\raggedright
12:30 -- 13:00\end{minipage}} & \multicolumn{1}{p{0.912in}|}{\begin{minipage}[t]{0.912in}\raggedright
Closure\end{minipage}} & \multicolumn{1}{p{2.300in}|}{\begin{minipage}[t]{2.300in}\raggedright
\end{minipage}}\\
\hline
\multicolumn{1}{|p{1.288in}|}{\begin{minipage}[t]{1.288in}\raggedright
13:00\end{minipage}} & \multicolumn{2}{p{3.212in}|}{\begin{minipage}[t]{3.212in}\raggedright
\textit{ Lunch}\end{minipage}}\\
\hline
\end{longtable}

\vfill 



%


%% file: participants/participants.tex

%





\titl{List of participants}

\begin{table}[ht]
\begin{center}
\footnotesize
\begin{tabular}{|ll|l|}
\hline
Family Name & Given Name & email\\
\hline
Bassi & Angelo & bassi@ts.infn.it \\
Belli & Pierluigi & pierluigi.belli@roma2.infn.it \\
Bruno & Angelo & bruno@sa.infn.it \\
Capolupo & Antonio & capolupo@sa.infn.it \\
Clozza & Alberto & Alberto.Clozza@lnf.infn.it \\
Curceanu & Catalina Oana & petrascu@lnf.infn.it \\
D'Ariano & Giacomo Mauro & dariano@unipv.it \\
Dakic & Borivoje & borivoje.dakic@unvie.ac.at \\
Dalla Brida & Mattia & mattia.dallabrida@gmail.com \\
De Martini & Francesco & francesco.demartini@uniroma1.it \\
Di Domenico & Antonio & antonio.didomenico@roma1.infn.it \\
Di Marco & Alessandro & alessandro.dimarco@roma2.infn.it \\
Diaz-Torres & Alexis & diaztorres@yahoo.com \\
Donadi & Sandro & donadi@ts.infn.it \\
Ferialdi & Luca & ferialdi@ts.infn.it \\
Floreanini & Roberto & florean@ts.infn.it \\
Fomenko & Kirill & kirill.fomenko@lngs.infn.it \\
Gerlich & Stefan & stefan.gerlich@univie.ac.at \\
Gouanere & Michel & micgou@wanadoo.fr \\
Hayrapetyan & Armen & armen@physi.uni-heidelberg.de \\
Hiesmayr & Beatrix C. & Beatrix.Hiesmayr@univie.ac.at \\
Jungmann & Kalus & jungmann@kvi.nl \\
Kiesel & Nikolai & nikolai.kiesel@univie.ac.at \\
Liang & Yeong-Cherng & yeongcherng.liang@unige.ch \\
Marton & Johann & johhan.marton@oeaw.ac.at \\
Mavromatos & Nikolaos & nikolaos.mavromatos@kcl.ac.uk \\
Mayburov & Sergey & mayburov@sci.lebedev.ru \\
Milotti & Edoardo & milotti@ts.infn.it \\
Pasqua & Antonio & toto.pasqua@gmail.com \\
Piacentini & Fabrizio & f.piacentini@inrim.it \\
Rauch & Helmut & rauch@ati.ac.at \\
Rauschenbeutel & Arno & arno.rauschenbeutel@ati.ac.at \\
Rizzo & Alessandro & alessandro.rizzo@lnf.infn.it \\
Scordo & Alessandro & alessandro.scordo@lnf.infn.it \\
Ursin & Rupert & rupert.ursin@univie.ac.at \\
Vacchini & Bassano & bassano.vacchini@mi.infn.it \\
Vona & Nicola & nocola\_vona@yahoo.it \\
Widmann & Eberhard & eberhard.widmann@oeaw.ac.at \\
Zanghi & Pierantonio & zanghi@ge.infn.it \\
Zavattini & Guido & guido.zavattini@unife.it \\
\hline
\end{tabular}
\end{center}
\end{table}




%% file: photo/photo.tex

 
 
 

\titl{Conference Photos}

\vspace{2cm}

\vspace{1.5cm}
\begin{figure}[htbp]
\begin{center}
\includegraphics[width=0.45\linewidth]{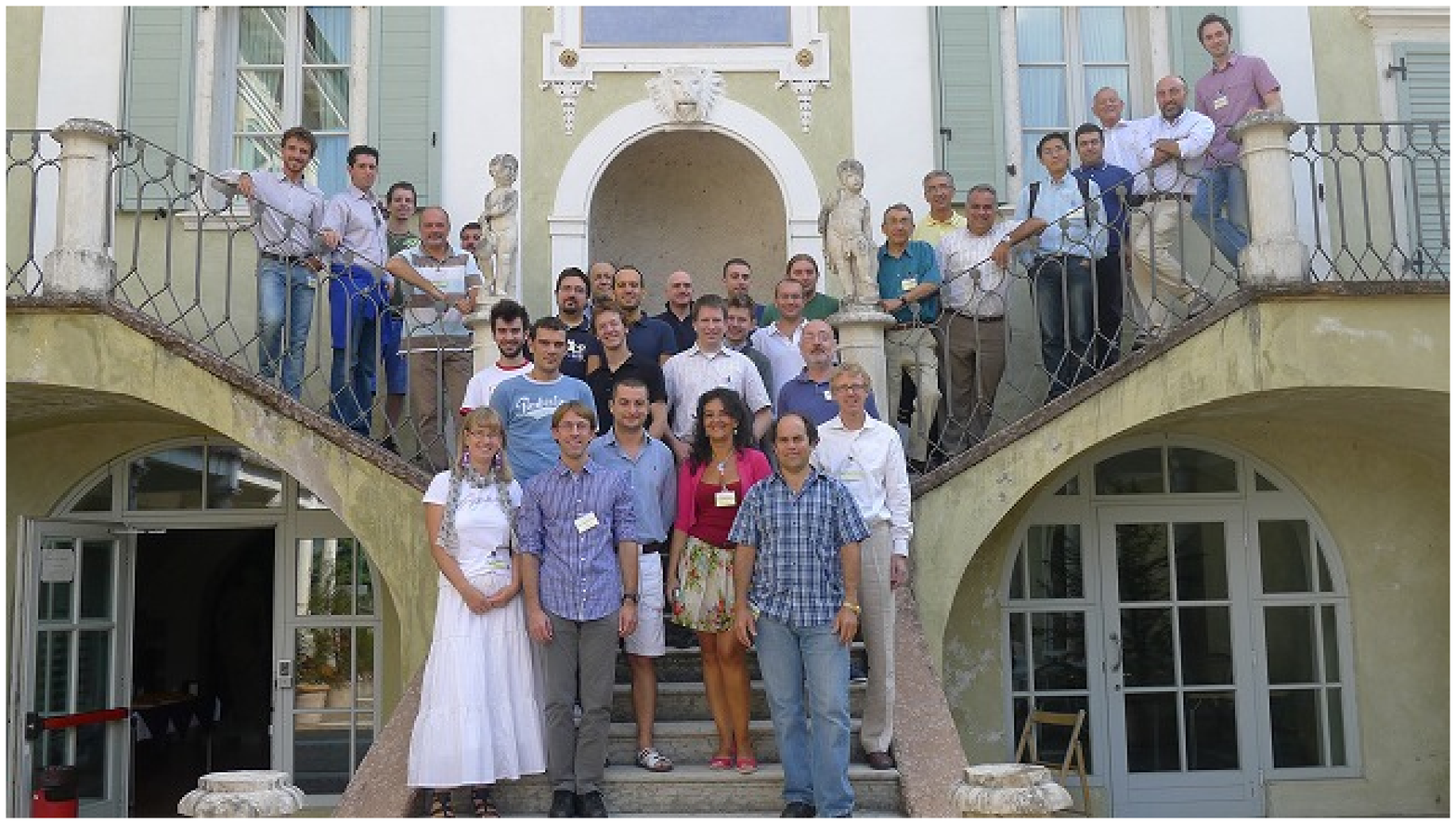}
\includegraphics[width=0.45\linewidth]{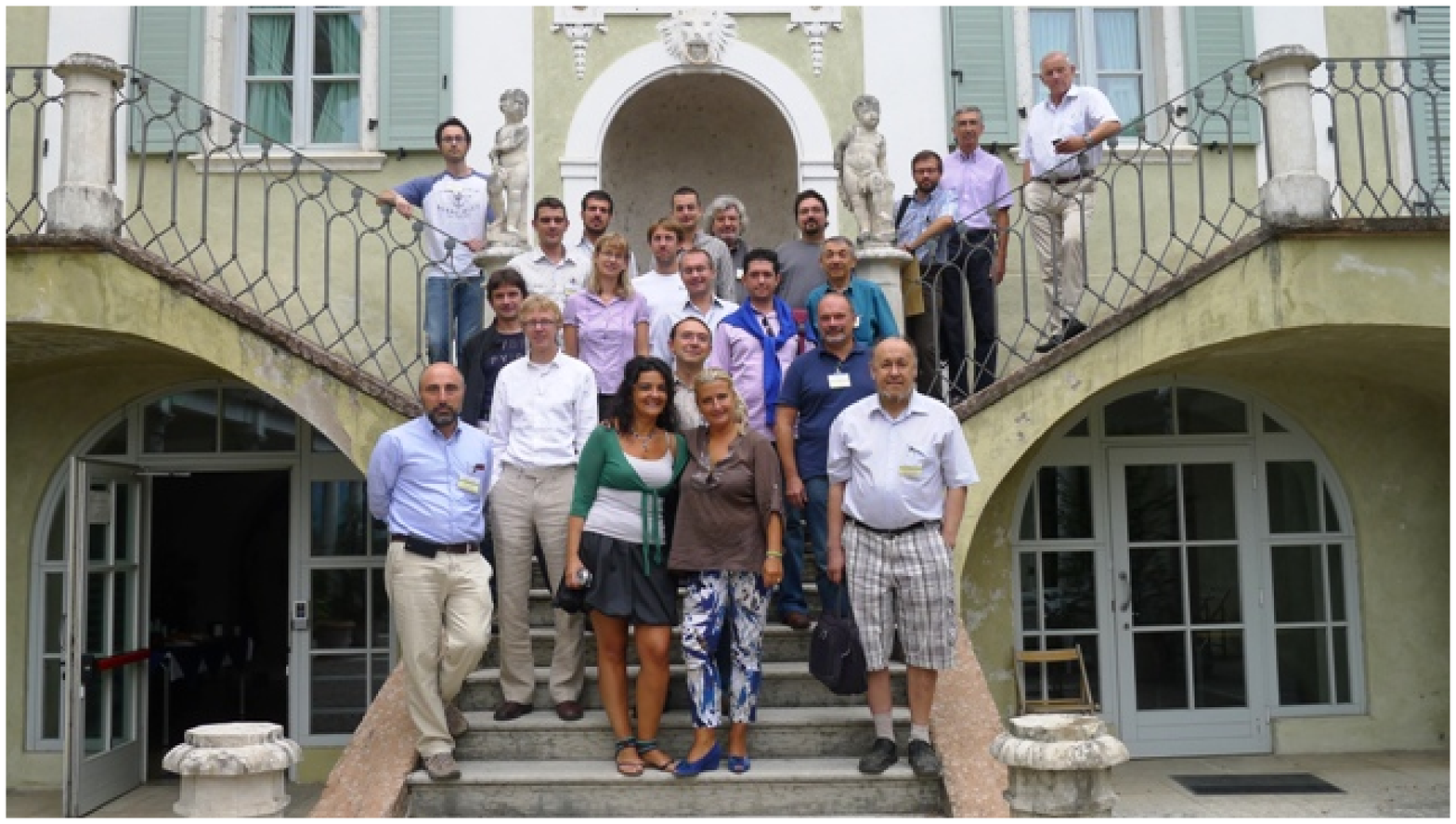}
\includegraphics[width=0.45\linewidth]{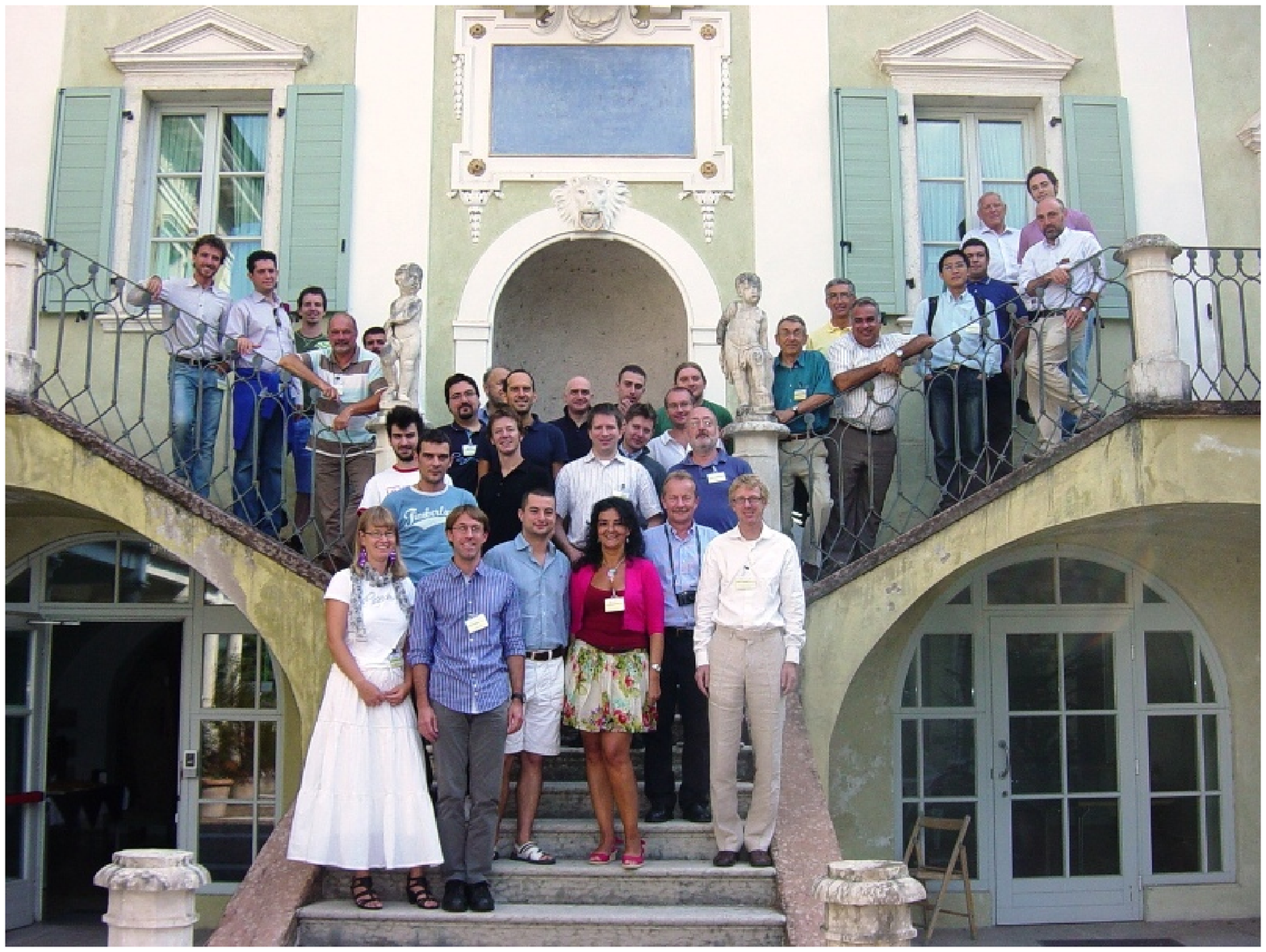}
\includegraphics[width=0.45\linewidth]{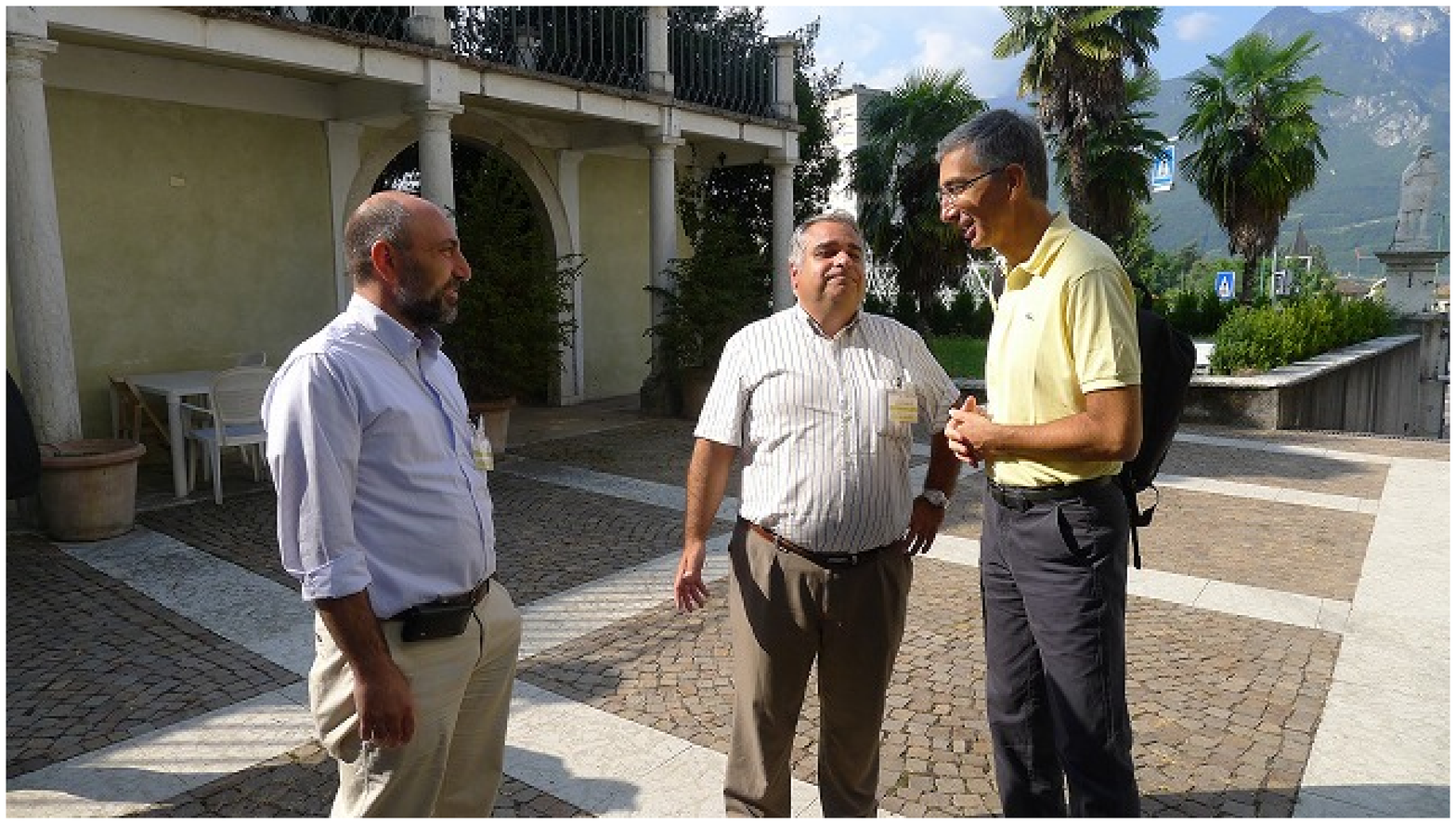}
\includegraphics[width=0.45\linewidth]{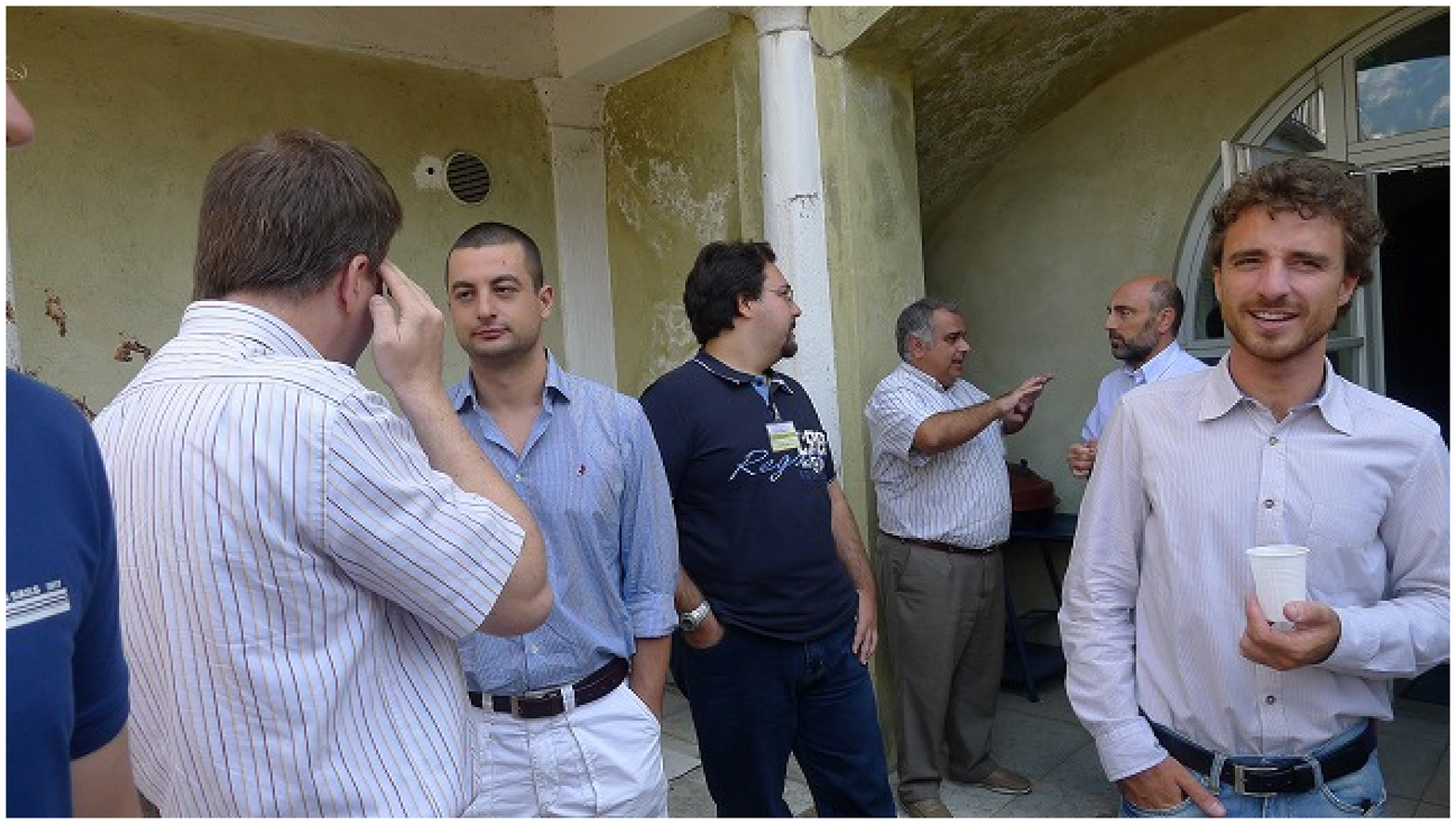}
\includegraphics[width=0.45\linewidth]{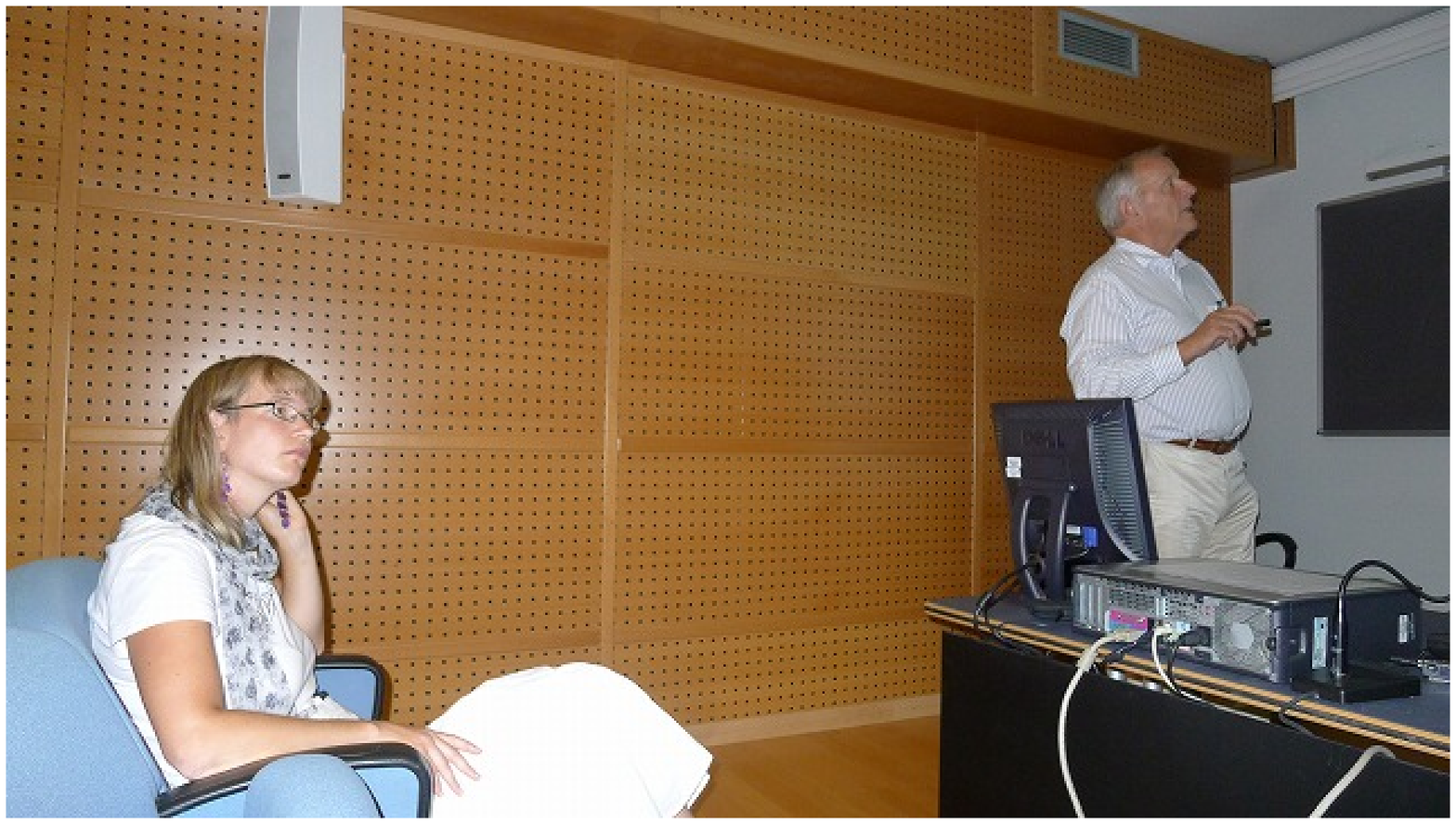}
\end{center}
\end{figure}
